\documentclass[aps,prx,twocolumn,nofootinbib,superscriptaddress,showpacs,showkeys,longbibliography,floatfix]{revtex4-2}

\usepackage{amsmath, amsxtra, amssymb, mathtools, isomath, nccmath, xspace, siunitx,graphicx}
\usepackage{color,hyperref,sidecap,natbib,dsfont,xcolor,pifont}
\usepackage[utf8]{inputenc}
\usepackage[british]{babel}
\usepackage{hhline}
\renewcommand{\vec}{\vectorsym}
\newcommand{\ket}[1]{\ensuremath{\lvert #1 \rangle}\xspace}%
\newcommand{\bra}[1]{\ensuremath{\langle #1 \rvert}\xspace}%
\renewcommand{\vec}{\vectorsym}
\sidecaptionvpos{figure}{h}
\long\def\symbolfootnote[#1]#2{\begingroup%
\def\thefootnote{\fnsymbol{footnote}}\footnotetext[#1]{#2}\endgroup}

\makeatletter
\@addtoreset{equation}{myequation}
\makeatother

\begin{document}
\setlength{\belowdisplayskip}{9pt}
\setlength{\abovedisplayskip}{9pt}
\setlength{\belowdisplayskip}{9pt}
\setlength{\abovedisplayskip}{9pt}

\title{\bf{Rydberg molecules bound by strong light fields}}


\author{Simon Hollerith}
\altaffiliation{These two authors contributed equally}
\affiliation{Max-Planck-Institut f\"{u}r Quantenoptik, 85748 Garching, Germany}
\affiliation{Munich Center for Quantum Science and Technology (MCQST), 80799 Munich, Germany}
\affiliation{Department of Physics, Harvard University, Cambridge, Massachusetts 02138, USA}

\author{Valentin Walther}
\altaffiliation{These two authors contributed equally}
\affiliation{Department of Physics, Harvard University, Cambridge, Massachusetts 02138, USA}
\affiliation{ITAMP, Harvard-Smithsonian Center for Astrophysics, Cambridge, Massachusetts 02138, USA}
\affiliation{ Department of Chemistry, Purdue University, West Lafayette, Indiana 47907, USA}
\affiliation{Department of Physics and Astronomy, Purdue University, West Lafayette, Indiana 47907, USA}

\author{Kritsana Srakaew}
\affiliation{Max-Planck-Institut f\"{u}r Quantenoptik, 85748 Garching, Germany}
\affiliation{Munich Center for Quantum Science and Technology (MCQST), 80799 Munich, Germany}

\author{David Wei}
\affiliation{Max-Planck-Institut f\"{u}r Quantenoptik, 85748 Garching, Germany}
\affiliation{Munich Center for Quantum Science and Technology (MCQST), 80799 Munich, Germany}

\author{Daniel Adler}
\affiliation{Max-Planck-Institut f\"{u}r Quantenoptik, 85748 Garching, Germany}
\affiliation{Munich Center for Quantum Science and Technology (MCQST), 80799 Munich, Germany}

\author{Suchita Agrawal}
\affiliation{Max-Planck-Institut f\"{u}r Quantenoptik, 85748 Garching, Germany}
\affiliation{Munich Center for Quantum Science and Technology (MCQST), 80799 Munich, Germany}

\author{Pascal Weckesser}
\affiliation{Max-Planck-Institut f\"{u}r Quantenoptik, 85748 Garching, Germany}
\affiliation{Munich Center for Quantum Science and Technology (MCQST), 80799 Munich, Germany}

\author{Immanuel Bloch}
\affiliation{Max-Planck-Institut f\"{u}r Quantenoptik, 85748 Garching, Germany}
\affiliation{Munich Center for Quantum Science and Technology (MCQST), 80799 Munich, Germany}
\affiliation{Fakultät für Physik, Ludwig-Maximilians-Universität München, 80799 München, Germany}

\author{Johannes Zeiher}
\affiliation{Max-Planck-Institut f\"{u}r Quantenoptik, 85748 Garching, Germany}
\affiliation{Munich Center for Quantum Science and Technology (MCQST), 80799 Munich, Germany}
\affiliation{Fakultät für Physik, Ludwig-Maximilians-Universität München, 80799 München, Germany}
\date{\today}

\begin{abstract}
The coupling of an isolated quantum state to a continuum is typically associated with decoherence and decreased lifetime. Here, we demonstrate that Rydberg macrodimers, weakly bound pairs of Rydberg atoms, can overcome this dissipative mechanism and instead form bound states with the continuum of free motional states. This is enabled by the unique combination of extraordinarily slow vibrational motion in the molecular state and the optical coupling to a non-interacting continuum. 
Under conditions of strong coupling, we observe the emergence of distinct resonances and explain them within a Fano model.
For atoms arranged on a lattice, we predict the strong continuum coupling to even stabilize molecules consisting of more than two atoms and find first signatures of these by observing atom loss correlations using a quantum gas microscope.
Our results present an intriguing mechanism to control decoherence and bind multiatomic molecules using strong light-matter interactions.
\end{abstract}

\maketitle

Coupling to a continuum of free states provides a prominent mechanism for quantum states to acquire a finite lifetime.
Examples include the decay of excited atoms via coupling to free electromagnetic modes, the phononic bath in molecules and solids, or quasiparticles coupled to a many-body continuum.
Most often, this coupling is weak and can be described perturbatively, giving rise to a finite lifetime as described by Wigner-Weisskopf theory.
Fano provided an exact description for the continuum coupling in the context of optical absorption, where eigenstates in isolated quantum systems are broadened into resonances~\cite{Fano_paper,Fano_RevMod}.
Instances where this generic paradigm fails include bound states in the continuum~\cite{Hsu2016,Longhi2007}, investigated accross various physical systems such as optical waveguides, molecules~\cite{Blech2020}, and electronic states in solids. Alternatively, in the less investigated regime of strong continuum couplings, stable eigenstates can emerge by shifting the bound state out of the coupled continuum~\cite{Verresen2019,Fano_strong_coupling,B_Gaveau_1995,strong_coupling_Sherman}.

\begin{figure*}
  \centering
  \includegraphics[width=1.\textwidth]{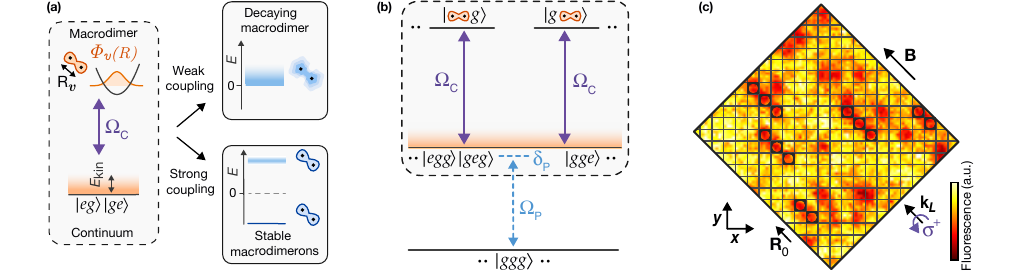}
  \caption{\label{fig:1}\label{fig:mac_block} \textbf{Tunable continuum coupling and system overview.} 
    \textbf{(a)} 
    Macrodimers (orange symbols) in a vibrational mode $\Phi_v(R)$ at a bond length $R_v$ are coupled into a continuum of free pair states (blurry orange region). 
    For weak couplings $\hbar\mathrm\Omega_\mathrm{C} \ll E_{\mathrm{kin}}$, smaller than the width of the coupled continuum, this broadens the macrodimer state (blurry blue region) and therefore reduces its lifetime. 
    For strong couplings $\hbar\mathrm\Omega_\mathrm{C} \gg E_{\mathrm{kin}}$, in contrast, we find a stable macrodimeron at negative energies, which is shifted out of the continuum, and a quasi-stable macrodimeron resonance at positive energies (blue pictograms). 
  \textbf{(b)} 
   We study this coupling mechanism for atoms arranged in a lattice with spacing $R_0$ close to $R_{v}$. Macrodimer excitations are resonantly coupled into non-interacting states containing ground state atoms $\ket{g}$ and a single Rydberg atom $\ket{e}$ (purple lines). 
    In this arrangement, where adjacent macrodimers are coupled via a two-photon transition, a strong continuum coupling can even stabilize bound states of more than two particles.
      The coupled system (dashed-bordered box) can be probed from the collective ground state using a weak probe field $\mathrm\Omega_\mathrm{p}$ (dashed blue line) at a detuning $\delta_\mathrm{p}$ to the singly-excited states.
  \textbf{(c)} In the experiment, these bound states were excited from a unity-filled optical lattice (gray grid). Macrodimer excitation occurred between pairs aligned along the diagonal distance $\mathbf{R}_0$ of the array, which is parallel to the magnetic field $\mathbf{B}$ and the propagation direction $\mathbf{k}_L$ of the $\sigma^+$-polarized excitation laser.  Atoms contributing to the binding leave the lattice (illustrated by thick-bordered sites), the remaining atoms were detected via fluorescence imaging. }
\end{figure*}

Here, we explore the continuous transition from the familiar weak continuum coupling to strong couplings using Rydberg macrodimers, see Fig.~\ref{fig:1}\,\textbf{(a)}. Macrodimers are giant molecules of two atoms, both excited to their Rydberg states~\cite{Boisseau2002,Overstreet2009,Sassmannshausen2016,Zhao_Cs_macrodimers,Hollerith_2019,Hollerith2023}. 
Held together by dipolar forces across distances of hundreds of nanometers, 
macrodimers exhibit a vibrational spectrum characteristic of diatomic molecules. 
For covalently bound molecules, coherent control over the bound state is limited to strong external fields such as available in pulsed lasers~\cite{Control_bondmaking,Light_induced_CI,L_ind_molpots}.
In contrast, the weak bonds of long-range Rydberg molecules~\cite{Shaffer2018,Dressed_ion_pair_states,Thomas2018} allow for external control by much weaker static fields~\cite{Overstreet2009,atom_ion_dynamics,booth_lonrange_strong_elDM}, radiofrequency waves~\cite{Petrosyan_MW_dressing,RF_photodiss_LRM}, Casimir forces~\cite{Casimir_Macrodiemrs}, or continuous-wave laser light. 
Macrodimers therefore offer an opportune handle to optically associate and dissociate them on timescales much faster than their vibrational period such that molecular bonds can be coherently controlled using laser light.

Our experiments were performed in a quantum gas microscope, where atom pairs on adjacent optical lattice sites can be excited into bound macrodimers. We laser couple the macrodimer state resonantly to non-interacting pair states composed of one Rydberg and one ground-state atom, see Fig.~\ref{fig:1}\,\textbf{(a,b)}.
Our spectroscopic probe reveals new resonances emerging in the presence of the strong light coupling.
Supported by our calculations, we interpret these resonances as hybridized states between macrodimers and the continuum 
and name them ``macrodimerons". Furthermore, the spatial arrangement in the lattice facilitates an exchange coupling of macrodimer states via a two-photon process, giving rise to bound states between three (``macrotrimerons") or even more atoms. Using site-resolved detection, we observe correlated two-atom as well as three-atom loss at the predicted energies.

\begin{figure}
  \centering
  \includegraphics[width=1.0\columnwidth]{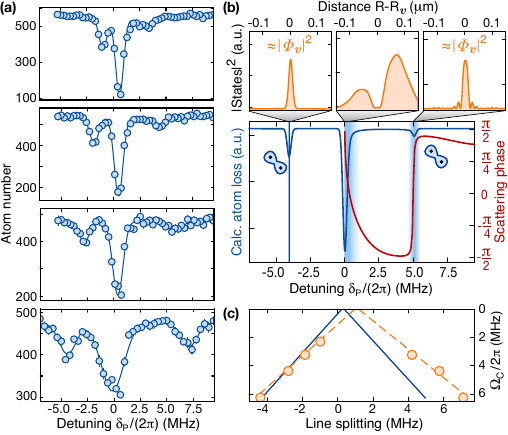}
  \caption{\label{fig:2}\textbf{Spectroscopic response and macrodimerons.} \textbf{(a)} Probe-field spectroscopy starting from an atomic array, with resonant coupling laser at Rabi frequencies $\mathrm\Omega_\mathrm{C}/2\pi =  \left[ 2.3,3.2,4.4, 6.2 \right] \,\si{\mega\hertz}$ (top to bottom).  
  All error bars denote one standard error of the mean (s.e.m.). \textbf{(b)} 
 The observed line splitting is already explained by a two-atom model which predicts a stable~(decaying) macrodimeron at negative~(positive) energies, here shown for $\mathrm\Omega_\mathrm{C}/2\pi =  6.2  \,\si{\mega\hertz}$, see Fig.~\ref{fig:1}\,\textbf{(a)}.
   The positive macrodimeron becomes quasi-stable at large couplings, featuring a narrow evolution of the associated scattering phase (red line).
  The spatial probability distributions of both macrodimeron types are similar to that of the macrodimer vibrational state $|\Phi_v (R)|^2$ (left/right subplot). The envelope of the central peak reflects the initial distribution $|\Phi_g (R)|^2$ of two atoms in the lattice ground state (central subplot). Its spatial components overlapping with the macrodimer are optically transferred to the macrodimeron states. The asymmetry of the resulting dip originates from a small mismatch between $R_0$ and $R_v$. 
  \textbf{(c)} The observed splitting between both macrodimerons (orange dashed line) is in reasonable agreement with \emph{ab initio} calculations including the electronic structure of the macrodimer (solid blue line). }
\end{figure}

The experiments started with a unity-filled two dimensional optical lattice of $^{87}\mathrm{Rb}$ atoms with lattice spacing $a_{\mathrm{lat}} = 532\,\si{\nano\meter}$~\cite{Hollerith_2019}, see Fig.~\ref{fig:1}\,\textbf{(c)}.
The atoms were initially prepared in the motional ground state at the individual lattice sites.
A $\sigma^+$-polarized coupling laser at a wavelength $\lambda = 298\,\si{\nano\meter}$ in the ultraviolet (UV) spectral range was propagating along the lattice diagonal direction, parallel to a magnetic bias field $B = 0.5\,\mathrm{G}$ setting the quantization axis.
The chosen $1_u$ macrodimer potential binds atom pairs at a separation $R_v = 712\,\si{\nano\meter}$, close to the lattice diagonal distance $R_0 \approx \sqrt{2}a_{\mathrm{lat}}$\,\cite{Hollerith_2021}.
The corresponding macrodimer state is $\ket{\Psi_{v}} = \Phi_{v}(R)\otimes \ket{\Psi_{\mathrm{el}}^{(2)}}$, with $\Phi_{v}(R)$ the lowest vibrational mode in the binding potential and $\ket{\Psi_{\mathrm{el}}^{(2)}}$ the electronic wave function.
For an orientation $\mathbf{R}_0 = (-1,1)a_{\mathrm{lat}}$ parallel to $\mathbf{B}$, the coupling laser couples $\ket{\Psi_{\mathrm{el}}^{(2)}}$ to non-interacting pair states $\ket{eg}$ and $\ket{ge}$ at a Rabi frequency $\mathrm\Omega_\mathrm{C}$, with the Rydberg state $\ket{e} = \ket{36P_{1/2},m_J=-1/2}$ and the hyperfine ground state $\ket{g} = \ket{F=2,m_F=-2}$.
In this configuration, excitation of macrodimers oriented orthogonal to $\mathbf{B}$ is suppressed and the lattice decouples into several one-dimensional systems~\cite{Hollerith_2021}.
We weakly phase-modulated our coupling laser tuned to a frequency close to the interaction shift $U = 735.2\,\si{\mega\hertz}$ between the macrodimer state and the pair state $\ket{ee}$ at large distances.
This enables probing the strongly coupled system via the transition $\ket{g}$ to $\ket{e} $ using a low-power probe field at a Rabi frequency $\mathrm\Omega_\mathrm{p}$ and a tunable detuning $\delta_\mathrm{p}$ using the red modulated sideband, see Fig.~\ref{fig:1}\,\textbf{(b)}.

To reveal the emergence of macrodimerons, we first performed spectroscopy for various coupling strengths $\mathrm\Omega_\mathrm{C}$. The macrodimer state was resonantly coupled to the singly-excited states by the coupling laser, whereas the weaker probe laser detuning was scanned.
We observed a splitting of the ground-state to macrodimer transition into three spectroscopically resolved lines, with a central resonance at a detuning $\delta_P = 0$ that remained unshifted and two outer resonances exhibiting a line shift that depends linearly on $\mathrm\Omega_\mathrm{C}$, see Fig.~\ref{fig:2}\,\textbf{(a)}. 
 While the resonance at negative detunings had a narrow and distinct line shape at all coupling strengths, the resonance at positive detunings was first broadened and, surprisingly, became more distinct at larger $\mathrm\Omega_\mathrm{C}$.

To understand the emergence of the observed spectrum, we first employ a generalized Fano-model involving two atoms, see Fig.~\ref{fig:2}\,\textbf{(b)}.
In the rotating frame, the Hamiltonian reads 
\begin{align}\label{eq:Fano_H}
\frac{\mathcal{H}}{\hbar} &= \int dk \varrho_k\omega_k  \Bigl( \ket{eg}\bra{eg} + \ket{ge}\bra{ge}  \Bigr)\ket{k} \bra{k} -\mathrm\Delta_\mathrm{C}   \ket{\Psi_{v}}\bra{\Psi_{v}} \notag \\
&+ \frac{\mathrm\Omega_\mathrm{C} }{2} \int dk \varrho_kf_k \Bigl( \ket{ge} + \ket{eg} \Bigr)\ket{k} \bra{\Psi_{v}} + \text{h.c.}, 
\end{align}
with plane wave modes $|k\rangle$ and their density of states $\varrho_k$ in the relative coordinate $R$ and the laser detuning $\mathrm\Delta_\mathrm{C}$ from the transition between the macrodimer and states $\ket{ge}$ and $\ket{eg}$ at rest ($k = 0$).
Contrary to the commonly used conditions of the Fano model, the continuum is bounded from below because of the exclusively positive kinetic energies $\hbar\omega_k$.
The optical coupling 
depends on $\mathrm\Omega_\mathrm{C}$ as well as the overlap integrals $f_k$ between  $\ket{k}$ and $\Phi_v(R)$. 
The macrodimer mostly couples to states $\ket{k}$ up to the kinetic energy $E_{\mathrm{kin}} = \int dk \varrho_k |f_k|^2\hbar\omega_k \approx h \times 850\,\si{\kilo\hertz} $ stored in its vibrational mode $\Phi_v(R)$.
The kinetic energy in the center-of-mass and angular coordinates, as well as the lattice potential, are much smaller than the vibrational energy and can be neglected.
We discretize the continuum and solve Eq.~\ref{eq:Fano_H} by exact diagonalization for resonant coupling $\mathrm\Delta_C = 0$.
 \begin{figure*}
  \centering
  \includegraphics[width=0.75\textwidth]{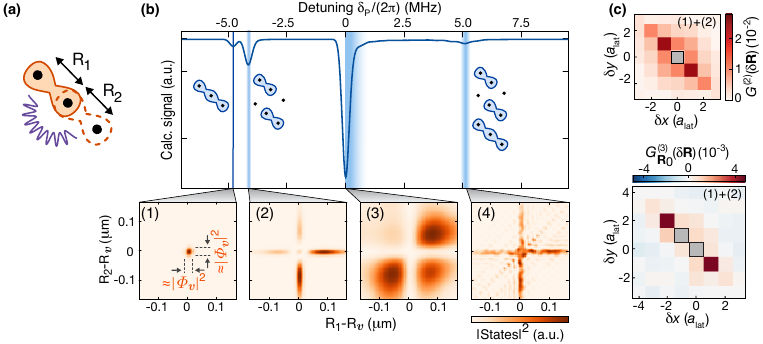}
  \caption{\label{fig:3}\textbf{Macrotrimerons and their microscopic signatures.} 
  \textbf{(a)} For three atoms with both relative distances, $R_1$ and $R_2$, close to the macrodimer bond length, the laser couples both adjacent macrodimers via a joint Rydberg excitation and its motional continuum. \textbf{(b)} 
 The calculated three-atom spectrum for $\mathrm\Omega_\mathrm{C}/2\pi =  6.2  \,\si{\mega\hertz}$ now exhibits motional states with contributions from two atom pairs (lower subplots). 
  At negative detunings, we identify a state (1), where $R_1$ and $R_2$ are confined to $\Phi_{v}(R)$, the macrotrimeron.
  The strong signal (2) at $\delta_\mathrm{p}<0$ originates from macrodimeron states where either the first or the second atom pair is confined to the vibrational mode, while the third atom is unconfined. 
  Again, we find an unshifted signature (3) covering motional states which have zero spatial overlap with $\Phi_v(R_1)$ as well as $\Phi_v(R_2)$ but are accessible from the broader three-atom ground state $\Phi_g(R_1)\Phi_g(R_2) \otimes \ket{ggg}$.
  At positive detunings, the continuum broadens the calculated contributions from two and three atoms into a single resonance (4). \textbf{(c)} Atom-loss correlations after illuminating the atoms for $50\si{\micro\second}$ with the laser tuned into resonance with the line observed at $\delta_\mathrm{p}<0$ where dimerons and trimerons are expected to contribute. Correlations $G^{(2)}(\delta\mathbf{R})$ and $G^{(3)}_{\mathbf{R}_0}(\delta\mathbf{R})$ show a two-atom and three-atom loss signal for atoms aligned along the expected lattice diagonal distance. At zero distances, the signals were excluded (gray). }
\end{figure*}

The optical transition strengths from the lattice ground state are provided by their overlaps with the obtained eigenstates, see Fig.~\ref{fig:2}\,\textbf{(b)}.
For negative $\delta_\mathrm{p} < 0$, we predict a signature originating from a single eigenstate $\ket{\bar\Psi^{(2)}_{v}}\approx  \Phi_{v}(R)\otimes \ket{\bar\Psi^{(2)}_{\mathrm{el}}}$, which we identify with the negative macrodimeron. Its electronic wave function $\ket{\bar\Psi_{\mathrm{el}}^{(2)}} = \frac{1}{\sqrt{2}}\ket{\Psi_{\mathrm{el}}^{(2)}} - \frac{1}{2}\left(\ket{ge} + \ket{eg}\right) $  is a superposition of the macrodimer and motionally unconfined singly-excited states. As a consequence, without the presence of the light field, an atom pair in the electronic state $\ket{\bar\Psi_{\mathrm{el}}^{(2)}}$ could not be stabilized in a bound motional state. However, the light coupling confines the motional state of the macrodimeron $\ket{\bar\Psi^{(2)}_{v}}$ to the vibrational mode $\Phi_{v}(R)$, resulting in a stable configuration. 
Contrary to the negative macrodimeron, the positive macrodimeron for $\delta_\mathrm{p} > 0$ is immersed in the continuum to which it resonantly couples.
Here, the description is equivalent to a scattering problem between an open continuum channel and a closed macrodimer channel~\cite{Feshbach_Cheng_2010}.
Only at large couplings $\hbar\mathrm\Omega_\mathrm{C} \gg E_{\mathrm{kin}}$, where the positive macrodimeron is shifted away from most of the coupled continuum states, we predict a narrow resonance. 
We infer that the positive macrodimeron is a quasi-bound state as we find a rapid evolution of the associated scattering phase across the resonance~\cite{taylor_scattering}.
The remaining broadening of the positive macrodimeron can be attributed to the much weaker coupling to high $k$-modes, which also results in a slightly modified, but still highly localized motional state, see upper right inset in Fig.~\ref{fig:2}\,\textbf{(b)}.
In this regime of strong couplings, where the motional state $\Phi_v(R)$ acts as a spectator to the strong coupling of the electronic states, the macrodimeron splitting is reminiscent of an Autler-Townes splitting. 
The central unshifted resonance at $\delta_P = 0$ arises because the relative wave function $\Phi_{g}(R)$ of two ground-state atoms in the lattice is much broader than the vibrational mode $\Phi_{v}(R)$. 
As a consequence, some modes $\ket{k}$ within the singly-excited manifold that are accessible from the two-atom ground state $\Phi_{g}(R)\otimes \ket{gg}$ but have zero overlap with $\Phi_{v}(R)$ remain uncoupled from the macrodimer state.
The relative signal strength between the two outer resonances and the central resonance depends on the overlap between $\Phi_{g}(R)$ and $\Phi_{v}(R)$ and can be tuned by the trap depth.
The deviation between the calculated and observed resonance position at $\delta_\mathrm{p}>0$ (see Fig.~\ref{fig:2}\,\textbf{(c)}) may originate from higher vibrational modes as well as perturbations within the macrodimer binding potential~\cite{SI,Weber2017}.

While the simplified two-atom Fano model 
captures the main spectroscopic signatures, it does not account for the lattice, where each state $\ket{e}$ is not coupled to one but to two adjacent macrodimers. 
Leading-order corrections are already expected for three equidistantly spaced atoms on a line, with electronic states $\ket{gge},\ket{geg},\ket{egg},\ket{\Psi^{(2)}_{\mathrm{el}}\, g}$ and $ \ket{g\,\Psi^{(2)}_{\mathrm{el}}}$ contributing to the Fano model, see Fig.~\ref{fig:1}\,\textbf{(b)}.
Here, one possible process is the ``hopping" of a macrodimer excitation between two adjacent atom pairs $\ket{\Psi^{(2)}_{\mathrm{el}}\, g}$ and $ \ket{g\,\Psi^{(2)}_{\mathrm{el}}}$, induced by the coupling laser through a resonant two-photon transition, see Fig.~\ref{fig:3}\,\textbf{(a)}.
For strong couplings, one expects the emergence of states where $\ket{\Psi^{(2)}_{\mathrm{el}}}$ is delocalized between the two pairs that can be formed from three atoms. An interesting question in this case relates to the nature of the motional states.
They can be parametrized in the two relative coordinates $R_1$ and $R_2$ of the first and the second atom pair, which we expand into plane waves $\ket{k_1}$ and $\ket{k_2}$. 
Interestingly, the model predicts similar spectra as the two-atom model, with loss signatures mainly occurring in three frequency regions, see Fig.~\ref{fig:3}\,\textbf{(b)}. 
Again, a central resonance combines singly-excited states which remain unshifted by the carrier field because their motional states have vanishing overlap with both vibrational macrodimer modes.
At negative detunings $\delta_\mathrm{p}$, we predict two closely spaced features which remained experimentally unresolved. The stronger of the two features includes two states where either the first or the second atom pair is bound into a macrodimeron, while the third atom remains unbound.
At slightly more negative $\delta_\mathrm{p}$, calculations predict the presence of yet another state $\ket{\bar\Psi^{(3)}_{v}} \approx \Phi_{v}(R_1)\,\Phi_{v}(R_2)\otimes \ket{\bar\Psi^{(3)}_{\mathrm{el}}}$, with an electronic wave function $\ket{\bar\Psi^{(3)}_{\mathrm{el}}} = -\frac{1}{2}\left( \ket{\Psi_{\mathrm{el}}^{(2)} g} + \ket{g \,\Psi_{\mathrm{el}}^{(2)}}\right) + \frac{1}{\sqrt{3}} \left( \frac{1}{2}\ket{gge} + \ket{geg} + \frac{1}{2}\ket{egg}\right) $.
We name this state, which contains components of both macrodimers, a macrotrimeron.
Remarkably, it adopts the motional state of the macrodimer in both relative coordinates and thus binds all three atoms, see Fig.~\ref{fig:3}\,\textbf{(b)}.
Although broadened by their immersion in the continuum, similar corresponding macrodimeron and -trimeron resonances appear for positive $\delta_\mathrm{p}$.



Even for the case where macrodimerons and -trimerons cannot be spectroscopically resolved, site-resolved atom loss correlations do allow to substantiate their presence.
Since the macrodimer hopping rate is much larger than the excitation rate by the optical probe and the present dissipation rates, we directly excite these quasiparticles as stable eigenstates of the light-matter Hamiltonian.
 The dominant binding energy is close to $ \mathrm\Omega_C$ and exceeds the lattice potential. Furthermore, the trapping potential of these molecules is deactivated by the contribution of antitrapped Rydberg states. As a consequence, all atoms bound by the light field will jointly disperse from their trap positions. 
Retrapping is even further suppressed by the large kinetic energy stored in the mode $\Phi_v(R)$, which will be eventually released after Rydberg decay.
Their excitation should therefore display fully correlated two-atom and three-atom loss signals. 
Measuring the two-atom loss correlation signal $G^{(2)}({\delta\mathbf{R}})$ from the reconstructed images at the location of the resonance at $\delta_\mathrm{p}<0$ shows the emergence of a macrodimeron signal at the expected distance $\delta\mathbf{R} = {\mathbf{R}_0}$, see Fig.~\ref{fig:3}\,\textbf{(c)}.
 Additionally, we evaluate three-atom loss correlations $G^{(3)}_{\mathbf{R}_0}(\delta\mathbf{R})$ (defined in~\cite{SI}).
 Conditioned on two atoms lost at a distance ${\mathbf{R}_0}$, we vary the separation $\delta\mathbf{R}$ to the third lost atom, see Fig.~\ref{fig:3}\,\textbf{(c)}.
 Indeed, as expected from macrotrimeron excitations, $G^{(3)}_{\mathbf{R}_0}(\delta\mathbf{R})$ shows a strong signal at $\delta\mathbf{R}  = (1,-1)\,a_{\mathrm{lat}}$ and $\delta\mathbf{R}  = (-2,2)\,a_{\mathrm{lat}}$ where all three lost atoms were aligned along the lattice diagonal direction.
 We extract a ratio of $3:1$ between two-atom and three-atom loss events by reproducing the observed correlations from numerically generated samples~\cite{SI}.
 \begin{figure}
  \centering
  \includegraphics[width=1.0\columnwidth]{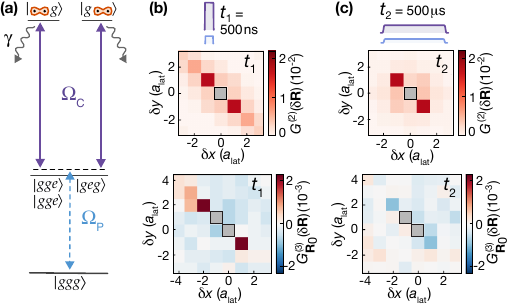}
  \caption{\label{fig:4}\textbf{Power dependence of the correlation signal.} \textbf{(a)} Macrodimers were resonantly excited from the ground state at varying coupling rates, 
 with slightly off-resonant singly-excited states~\cite{SI}.
\textbf{(b)} At high laser powers and large coupling rates $\mathrm\Omega_\mathrm{C}/2\pi = 7.8(3)\,\si{\mega\hertz}$, where the coupling between adjacent macrodimers exceeded the theoretical macrodimer decay rate $\gamma \approx 1/20\si{\micro\second}$, we observed two-atom loss correlations $G^{(2)}(\mathbf{R}_0)$ as well as three-atom loss correlations $G^{(3)}_{\mathbf{R}_0}(-\mathbf{R}_0)$ after a pulse time $t_1 = 500\,\si{\nano\second}$.
\textbf{(c)} In the opposite limit where $\gamma$ dominates, we predict to only excite isolated macrodimers. As expected, after a much longer pulse time $t_2 = 500\,\si{\micro\second}$, we only observed a two-atom loss signal. The observed negative signal at low powers can be understood from uncorrelated pair losses alone.}
  \end{figure}

To further substantiate the presence of macrotrimerons, we compared the regime where adjacent macrodimers are strongly coupled, with the opposite limit of weak couplings, where the intrinsic macrodimer decay rate dominates.
In the former case, we observed significant three-atom loss already at an illumination time of only $t_1 = 500\,\si{\nano\second}$, see Fig.~\ref{fig:4}\,\textbf{(a)}.
In the latter case, after a much longer pulse time $t_2 = 500\,\si{\micro\second}$ required to reach the same two-atom loss signal as before, we only observed pair losses but no three-atom loss. 
These observations confirm that the three-atom loss mechanism is directly associated with the high power UV pulse in the strongly coupled regime. 
Furthermore, these observations exclude a collisional loss model, whereby excited macrodimers are expelled from the lattice and subsequently knock out neighboring atoms because the pulse time $t_1$ is too short for the atoms to move from their sites.
Additional observations presented in~\cite{SI} exclude other alternative three-atom loss mechanisms, such as mechanisms based on radiative decay or Rydberg antiblockade~\cite{Simonelli_2016,Goldschmidt_2016}, but are consistent with our interpretation of light-induced macrodimer hopping.

In conclusion, we studied the light-coupling of Rydberg macrodimers to a continuum of free states. 
At strong couplings, we demonstrated that the coupling gives rise to new emerging quasiparticles that split off from the continuum and therefore are stable eigenstates. This can be seen as a reversal of the original idea of Fano, which endows a finite lifetime to an isolated quantum state. 
For systems beyond two atoms, we predict this mechanism to also stabilize macrotrimerons, complex quasiparticles with the motional state of a triatomic molecule, although the contributing Rydberg interactions are only binary. 
We observed microscopic three-atom loss signatures in a regime where such macrotrimerons are expected and further tested the presence of the underlying macrodimer hopping process by varying laser power and detuning to the singly-excited state.
The described process sets it apart from pure electronic excitation hopping \cite{symmetry_p_phase_2019,Chew2022,Facilitation_Antoine,kim2023realization} by the contribution of the motional mode of the macrodimer.
Such an interplay between electronic states and motional modes, recently also discussed in Rydberg facilitation models~\cite{kin_const_facilit,Matteo_jahn_teller}, opens up opportunities to construct molecular states of even larger particle number. 
First signatures were already found in some of our correlation signals, which indeed showed correlated four-atom events, see Fig.~\ref{fig:S1} in~\cite{SI} and~\cite{Hollerith2023}.
For our parameters, the associated excitation rates decrease by about one order of magnitude with each additional contributing atom because of the decreasing spatial overlap of the lattice ground state and the multi-atom bound state. 
In the future, this overlap could be improved using shallower binding potentials at higher principal quantum numbers where motional ground states can be prepared that are identical with the vibrational mode, providing much larger multi-atom contributions to the spectra~\cite{Hollerith2023}.
Finally, macrodimer hopping can also be studied in the time domain, e.g., by quenching the laser coupling and tracking the combined evolution of the electronic and motional states. 

\emph{Correspondence address: } Shollerith@fas.harvard.edu

\textbf{Acknowledgements: }
The authors thank Jun Rui, Johannes Deiglmayr, Philip Osterholz, Matteo Magoni, Alexander Schuckert, Elmer Guardado-Sanchez, Tijs Karman, and Simon Evered for discussions.
We acknowledge funding by the Max Planck Society (MPG) and the Deutsche Forschungsgemeinschaft (DFG, German Research Foundation) under Germany’s Excellence Strategy – EXC-2111–390814868. This publication has also received funding under Horizon Europe programme HORIZON-CL4-2022-QUANTUM-02-SGA via the project 101113690 (PASQuanS2.1). K. S. and S. A. acknowledge funding from the International Max Planck Research School (IMPRS) for Quantum Science and Technology.
J.Z. acknowledges support from the BMBF through the program “Quantum technologies - from basic research to market” (Grant No. 13N16265).
V.W. acknowledges support by the NSF through a grant for the Institute for Theoretical Atomic, Molecular, and Optical Physics at Harvard University and the Smithsonian Astrophysical Observatory.
S.H. acknowledges funding through the Harvard Quantum Initiative Postdoctoral Fellowship in Quantum Science and Engineering.
P.W. acknowledges funding through the Walter Benjamin programme (DFG project 516136618).

\bibliography{Dimerons_and_Trimerons}

\begin{thebibliography}{41}%
\makeatletter
\providecommand \@ifxundefined [1]{%
 \@ifx{#1\undefined}
}%
\providecommand \@ifnum [1]{%
 \ifnum #1\expandafter \@firstoftwo
 \else \expandafter \@secondoftwo
 \fi
}%
\providecommand \@ifx [1]{%
 \ifx #1\expandafter \@firstoftwo
 \else \expandafter \@secondoftwo
 \fi
}%
\providecommand \natexlab [1]{#1}%
\providecommand \enquote  [1]{``#1''}%
\providecommand \bibnamefont  [1]{#1}%
\providecommand \bibfnamefont [1]{#1}%
\providecommand \citenamefont [1]{#1}%
\providecommand \href@noop [0]{\@secondoftwo}%
\providecommand \href [0]{\begingroup \@sanitize@url \@href}%
\providecommand \@href[1]{\@@startlink{#1}\@@href}%
\providecommand \@@href[1]{\endgroup#1\@@endlink}%
\providecommand \@sanitize@url [0]{\catcode `\\12\catcode `\$12\catcode
  `\&12\catcode `\#12\catcode `\^12\catcode `\_12\catcode `\%12\relax}%
\providecommand \@@startlink[1]{}%
\providecommand \@@endlink[0]{}%
\providecommand \url  [0]{\begingroup\@sanitize@url \@url }%
\providecommand \@url [1]{\endgroup\@href {#1}{\urlprefix }}%
\providecommand \urlprefix  [0]{URL }%
\providecommand \Eprint [0]{\href }%
\providecommand \doibase [0]{https://doi.org/}%
\providecommand \selectlanguage [0]{\@gobble}%
\providecommand \bibinfo  [0]{\@secondoftwo}%
\providecommand \bibfield  [0]{\@secondoftwo}%
\providecommand \translation [1]{[#1]}%
\providecommand \BibitemOpen [0]{}%
\providecommand \bibitemStop [0]{}%
\providecommand \bibitemNoStop [0]{.\EOS\space}%
\providecommand \EOS [0]{\spacefactor3000\relax}%
\providecommand \BibitemShut  [1]{\csname bibitem#1\endcsname}%
\let\auto@bib@innerbib\@empty
\bibitem [{\citenamefont {Fano}(1961)}]{Fano_paper}%
  \BibitemOpen
  \bibfield  {author} {\bibinfo {author} {\bibfnamefont {U.}~\bibnamefont
  {Fano}},\ }\bibfield  {title} {\bibinfo {title} {{Effects of Configuration
  Interaction on Intensities and Phase Shifts}},\ }\href
  {https://doi.org/10.1103/PhysRev.124.1866} {\bibfield  {journal} {\bibinfo
  {journal} {Phys. Rev.}\ }\textbf {\bibinfo {volume} {124}},\ \bibinfo {pages}
  {1866} (\bibinfo {year} {1961})}\BibitemShut {NoStop}%
\bibitem [{\citenamefont {Miroshnichenko}\ \emph {et~al.}(2010)\citenamefont
  {Miroshnichenko}, \citenamefont {Flach},\ and\ \citenamefont
  {Kivshar}}]{Fano_RevMod}%
  \BibitemOpen
  \bibfield  {author} {\bibinfo {author} {\bibfnamefont {A.~E.}\ \bibnamefont
  {Miroshnichenko}}, \bibinfo {author} {\bibfnamefont {S.}~\bibnamefont
  {Flach}},\ and\ \bibinfo {author} {\bibfnamefont {Y.~S.}\ \bibnamefont
  {Kivshar}},\ }\bibfield  {title} {\bibinfo {title} {Fano resonances in
  nanoscale structures},\ }\href {https://doi.org/10.1103/RevModPhys.82.2257}
  {\bibfield  {journal} {\bibinfo  {journal} {Rev. Mod. Phys.}\ }\textbf
  {\bibinfo {volume} {82}},\ \bibinfo {pages} {2257} (\bibinfo {year}
  {2010})}\BibitemShut {NoStop}%
\bibitem [{\citenamefont {Hsu}\ \emph {et~al.}(2016)\citenamefont {Hsu},
  \citenamefont {Zhen}, \citenamefont {Stone}, \citenamefont {Joannopoulos},\
  and\ \citenamefont {Solja{\v{c}}i{\'{c}}}}]{Hsu2016}%
  \BibitemOpen
  \bibfield  {author} {\bibinfo {author} {\bibfnamefont {C.~W.}\ \bibnamefont
  {Hsu}}, \bibinfo {author} {\bibfnamefont {B.}~\bibnamefont {Zhen}}, \bibinfo
  {author} {\bibfnamefont {A.~D.}\ \bibnamefont {Stone}}, \bibinfo {author}
  {\bibfnamefont {J.~D.}\ \bibnamefont {Joannopoulos}},\ and\ \bibinfo {author}
  {\bibfnamefont {M.}~\bibnamefont {Solja{\v{c}}i{\'{c}}}},\ }\bibfield
  {title} {\bibinfo {title} {Bound states in the continuum},\ }\href
  {https://doi.org/10.1038/natrevmats.2016.48} {\bibfield  {journal} {\bibinfo
  {journal} {Nat. Rev. Mater.}\ }\textbf {\bibinfo {volume} {1}},\ \bibinfo
  {pages} {16048} (\bibinfo {year} {2016})}\BibitemShut {NoStop}%
\bibitem [{\citenamefont {Longhi}(2007)}]{Longhi2007}%
  \BibitemOpen
  \bibfield  {author} {\bibinfo {author} {\bibfnamefont {S.}~\bibnamefont
  {Longhi}},\ }\bibfield  {title} {\bibinfo {title} {{Bound states in the
  continuum in a single-level Fano-Anderson model}},\ }\href
  {https://doi.org/10.1140/epjb/e2007-00143-2} {\bibfield  {journal} {\bibinfo
  {journal} {Eur. Phys. J. B}\ }\textbf {\bibinfo {volume} {57}},\ \bibinfo
  {pages} {45} (\bibinfo {year} {2007})}\BibitemShut {NoStop}%
\bibitem [{\citenamefont {Blech}\ \emph {et~al.}(2020)\citenamefont {Blech},
  \citenamefont {Shagam}, \citenamefont {H{\"o}lsch}, \citenamefont {Paliwal},
  \citenamefont {Skomorowski}, \citenamefont {Rosenberg}, \citenamefont
  {Bibelnik}, \citenamefont {Heber}, \citenamefont {Reich}, \citenamefont
  {Narevicius},\ and\ \citenamefont {Koch}}]{Blech2020}%
  \BibitemOpen
  \bibfield  {author} {\bibinfo {author} {\bibfnamefont {A.}~\bibnamefont
  {Blech}}, \bibinfo {author} {\bibfnamefont {Y.}~\bibnamefont {Shagam}},
  \bibinfo {author} {\bibfnamefont {N.}~\bibnamefont {H{\"o}lsch}}, \bibinfo
  {author} {\bibfnamefont {P.}~\bibnamefont {Paliwal}}, \bibinfo {author}
  {\bibfnamefont {W.}~\bibnamefont {Skomorowski}}, \bibinfo {author}
  {\bibfnamefont {J.~W.}\ \bibnamefont {Rosenberg}}, \bibinfo {author}
  {\bibfnamefont {N.}~\bibnamefont {Bibelnik}}, \bibinfo {author}
  {\bibfnamefont {O.}~\bibnamefont {Heber}}, \bibinfo {author} {\bibfnamefont
  {D.~M.}\ \bibnamefont {Reich}}, \bibinfo {author} {\bibfnamefont
  {E.}~\bibnamefont {Narevicius}},\ and\ \bibinfo {author} {\bibfnamefont
  {C.~P.}\ \bibnamefont {Koch}},\ }\bibfield  {title} {\bibinfo {title} {{Phase
  protection of Fano-Feshbach resonances}},\ }\href
  {https://doi.org/10.1038/s41467-020-14797-w} {\bibfield  {journal} {\bibinfo
  {journal} {Nat. Commun.}\ }\textbf {\bibinfo {volume} {11}},\ \bibinfo
  {pages} {999} (\bibinfo {year} {2020})}\BibitemShut {NoStop}%
\bibitem [{\citenamefont {Verresen}\ \emph {et~al.}(2019)\citenamefont
  {Verresen}, \citenamefont {Moessner},\ and\ \citenamefont
  {Pollmann}}]{Verresen2019}%
  \BibitemOpen
  \bibfield  {author} {\bibinfo {author} {\bibfnamefont {R.}~\bibnamefont
  {Verresen}}, \bibinfo {author} {\bibfnamefont {R.}~\bibnamefont {Moessner}},\
  and\ \bibinfo {author} {\bibfnamefont {F.}~\bibnamefont {Pollmann}},\
  }\bibfield  {title} {\bibinfo {title} {Avoided quasiparticle decay from
  strong quantum interactions},\ }\href
  {https://doi.org/10.1038/s41567-019-0535-3} {\bibfield  {journal} {\bibinfo
  {journal} {Nat. Phys.}\ }\textbf {\bibinfo {volume} {15}},\ \bibinfo {pages}
  {750} (\bibinfo {year} {2019})}\BibitemShut {NoStop}%
\bibitem [{\citenamefont {Zeb}(2022)}]{Fano_strong_coupling}%
  \BibitemOpen
  \bibfield  {author} {\bibinfo {author} {\bibfnamefont {M.~A.}\ \bibnamefont
  {Zeb}},\ }\bibfield  {title} {\bibinfo {title} {Fano resonance in the
  strong-coupling regime},\ }\href
  {https://doi.org/10.1103/PhysRevB.106.155134} {\bibfield  {journal} {\bibinfo
   {journal} {Phys. Rev. B}\ }\textbf {\bibinfo {volume} {106}},\ \bibinfo
  {pages} {155134} (\bibinfo {year} {2022})}\BibitemShut {NoStop}%
\bibitem [{\citenamefont {Gaveau}\ and\ \citenamefont
  {Schulman}(1995)}]{B_Gaveau_1995}%
  \BibitemOpen
  \bibfield  {author} {\bibinfo {author} {\bibfnamefont {B.}~\bibnamefont
  {Gaveau}}\ and\ \bibinfo {author} {\bibfnamefont {L.~S.}\ \bibnamefont
  {Schulman}},\ }\bibfield  {title} {\bibinfo {title} {Limited quantum decay},\
  }\href {https://doi.org/10.1088/0305-4470/28/24/029} {\bibfield  {journal}
  {\bibinfo  {journal} {J. Phys. A Math. Gen}\ }\textbf {\bibinfo {volume}
  {28}},\ \bibinfo {pages} {7359} (\bibinfo {year} {1995})}\BibitemShut
  {NoStop}%
\bibitem [{\citenamefont {Kofman}\ \emph {et~al.}(1994)\citenamefont {Kofman},
  \citenamefont {Kurizki},\ and\ \citenamefont
  {Sherman}}]{strong_coupling_Sherman}%
  \BibitemOpen
  \bibfield  {author} {\bibinfo {author} {\bibfnamefont {A.~G.}\ \bibnamefont
  {Kofman}}, \bibinfo {author} {\bibfnamefont {G.}~\bibnamefont {Kurizki}},\
  and\ \bibinfo {author} {\bibfnamefont {B.}~\bibnamefont {Sherman}},\
  }\bibfield  {title} {\bibinfo {title} {{Spontaneous and Induced Atomic Decay
  in Photonic Band Structures}},\ }\href
  {https://doi.org/10.1080/09500349414550381} {\bibfield  {journal} {\bibinfo
  {journal} {J. Mod. Opt.}\ }\textbf {\bibinfo {volume} {41}},\ \bibinfo
  {pages} {353} (\bibinfo {year} {1994})}\BibitemShut {NoStop}%
\bibitem [{\citenamefont {Boisseau}\ \emph {et~al.}(2002)\citenamefont
  {Boisseau}, \citenamefont {Simbotin},\ and\ \citenamefont
  {C\^ot\'e}}]{Boisseau2002}%
  \BibitemOpen
  \bibfield  {author} {\bibinfo {author} {\bibfnamefont {C.}~\bibnamefont
  {Boisseau}}, \bibinfo {author} {\bibfnamefont {I.}~\bibnamefont {Simbotin}},\
  and\ \bibinfo {author} {\bibfnamefont {R.}~\bibnamefont {C\^ot\'e}},\
  }\bibfield  {title} {\bibinfo {title} {{Macrodimers: Ultralong Range Rydberg
  Molecules}},\ }\href {https://doi.org/10.1103/PhysRevLett.88.133004}
  {\bibfield  {journal} {\bibinfo  {journal} {Phys. Rev. Lett.}\ }\textbf
  {\bibinfo {volume} {88}},\ \bibinfo {pages} {133004} (\bibinfo {year}
  {2002})}\BibitemShut {NoStop}%
\bibitem [{\citenamefont {Overstreet}\ \emph {et~al.}(2009)\citenamefont
  {Overstreet}, \citenamefont {Schwettmann}, \citenamefont {Tallant},
  \citenamefont {Booth},\ and\ \citenamefont {Shaffer}}]{Overstreet2009}%
  \BibitemOpen
  \bibfield  {author} {\bibinfo {author} {\bibfnamefont {K.~R.}\ \bibnamefont
  {Overstreet}}, \bibinfo {author} {\bibfnamefont {A.}~\bibnamefont
  {Schwettmann}}, \bibinfo {author} {\bibfnamefont {J.}~\bibnamefont
  {Tallant}}, \bibinfo {author} {\bibfnamefont {D.}~\bibnamefont {Booth}},\
  and\ \bibinfo {author} {\bibfnamefont {J.~P.}\ \bibnamefont {Shaffer}},\
  }\bibfield  {title} {\bibinfo {title} {{Observation of Electric-Field-Induced
  {{Cs Rydberg}} Atom Macrodimers}},\ }\href
  {https://doi.org/10.1038/nphys1307} {\bibfield  {journal} {\bibinfo
  {journal} {Nat. Phys.}\ }\textbf {\bibinfo {volume} {5}},\ \bibinfo {pages}
  {581} (\bibinfo {year} {2009})}\BibitemShut {NoStop}%
\bibitem [{\citenamefont {Sa{\ss}mannshausen}\ and\ \citenamefont
  {Deiglmayr}(2016)}]{Sassmannshausen2016}%
  \BibitemOpen
  \bibfield  {author} {\bibinfo {author} {\bibfnamefont {H.}~\bibnamefont
  {Sa{\ss}mannshausen}}\ and\ \bibinfo {author} {\bibfnamefont
  {J.}~\bibnamefont {Deiglmayr}},\ }\bibfield  {title} {\bibinfo {title}
  {{Observation of Rydberg-Atom Macrodimers: Micrometer-Sized Diatomic
  Molecules}},\ }\href {https://doi.org/10.1103/PhysRevLett.117.083401}
  {\bibfield  {journal} {\bibinfo  {journal} {Phys. Rev. Lett.}\ }\textbf
  {\bibinfo {volume} {117}},\ \bibinfo {pages} {083401} (\bibinfo {year}
  {2016})}\BibitemShut {NoStop}%
\bibitem [{\citenamefont {Bai}\ \emph {et~al.}(2023)\citenamefont {Bai},
  \citenamefont {Jiao}, \citenamefont {Song}, \citenamefont {Raithel},
  \citenamefont {Jia},\ and\ \citenamefont {Zhao}}]{Zhao_Cs_macrodimers}%
  \BibitemOpen
  \bibfield  {author} {\bibinfo {author} {\bibfnamefont {J.}~\bibnamefont
  {Bai}}, \bibinfo {author} {\bibfnamefont {Y.}~\bibnamefont {Jiao}}, \bibinfo
  {author} {\bibfnamefont {R.}~\bibnamefont {Song}}, \bibinfo {author}
  {\bibfnamefont {G.}~\bibnamefont {Raithel}}, \bibinfo {author} {\bibfnamefont
  {S.}~\bibnamefont {Jia}},\ and\ \bibinfo {author} {\bibfnamefont
  {J.}~\bibnamefont {Zhao}},\ }\href@noop {} {\bibinfo {title} {{Microwave
  photo-association of fine-structure-induced Rydberg $(n+2)D_{5/2}nF_{J}$
  macro-dimer molecules of cesium}}} (\bibinfo {year} {2023}),\ \Eprint
  {https://arxiv.org/abs/2310.11934} {arXiv:2310.11934} \BibitemShut {NoStop}%
\bibitem [{\citenamefont {Hollerith}\ \emph {et~al.}(2019)\citenamefont
  {Hollerith}, \citenamefont {Zeiher}, \citenamefont {Rui}, \citenamefont
  {Rubio-Abadal}, \citenamefont {Walther}, \citenamefont {Pohl}, \citenamefont
  {Stamper-Kurn}, \citenamefont {Bloch},\ and\ \citenamefont
  {Gross}}]{Hollerith_2019}%
  \BibitemOpen
  \bibfield  {author} {\bibinfo {author} {\bibfnamefont {S.}~\bibnamefont
  {Hollerith}}, \bibinfo {author} {\bibfnamefont {J.}~\bibnamefont {Zeiher}},
  \bibinfo {author} {\bibfnamefont {J.}~\bibnamefont {Rui}}, \bibinfo {author}
  {\bibfnamefont {A.}~\bibnamefont {Rubio-Abadal}}, \bibinfo {author}
  {\bibfnamefont {V.}~\bibnamefont {Walther}}, \bibinfo {author} {\bibfnamefont
  {T.}~\bibnamefont {Pohl}}, \bibinfo {author} {\bibfnamefont {D.~M.}\
  \bibnamefont {Stamper-Kurn}}, \bibinfo {author} {\bibfnamefont
  {I.}~\bibnamefont {Bloch}},\ and\ \bibinfo {author} {\bibfnamefont
  {C.}~\bibnamefont {Gross}},\ }\bibfield  {title} {\bibinfo {title} {{Quantum
  gas microscopy of Rydberg macrodimers}},\ }\href
  {https://doi.org/10.1126/science.aaw4150} {\bibfield  {journal} {\bibinfo
  {journal} {Science}\ }\textbf {\bibinfo {volume} {364}},\ \bibinfo {pages}
  {664} (\bibinfo {year} {2019})}\BibitemShut {NoStop}%
\bibitem [{\citenamefont {Hollerith}\ and\ \citenamefont
  {Zeiher}(2023)}]{Hollerith2023}%
  \BibitemOpen
  \bibfield  {author} {\bibinfo {author} {\bibfnamefont {S.}~\bibnamefont
  {Hollerith}}\ and\ \bibinfo {author} {\bibfnamefont {J.}~\bibnamefont
  {Zeiher}},\ }\bibfield  {title} {\bibinfo {title} {Rydberg macrodimers:
  Diatomic molecules on the micrometer scale},\ }\href
  {https://doi.org/10.1021/acs.jpca.2c08454} {\bibfield  {journal} {\bibinfo
  {journal} {J. Phys. Chem. A}\ }\textbf {\bibinfo {volume} {127}},\ \bibinfo
  {pages} {3925} (\bibinfo {year} {2023})}\BibitemShut {NoStop}%
\bibitem [{\citenamefont {Levin}\ \emph {et~al.}(2015)\citenamefont {Levin},
  \citenamefont {Skomorowski}, \citenamefont {Rybak}, \citenamefont {Kosloff},
  \citenamefont {Koch},\ and\ \citenamefont {Amitay}}]{Control_bondmaking}%
  \BibitemOpen
  \bibfield  {author} {\bibinfo {author} {\bibfnamefont {L.}~\bibnamefont
  {Levin}}, \bibinfo {author} {\bibfnamefont {W.}~\bibnamefont {Skomorowski}},
  \bibinfo {author} {\bibfnamefont {L.}~\bibnamefont {Rybak}}, \bibinfo
  {author} {\bibfnamefont {R.}~\bibnamefont {Kosloff}}, \bibinfo {author}
  {\bibfnamefont {C.~P.}\ \bibnamefont {Koch}},\ and\ \bibinfo {author}
  {\bibfnamefont {Z.}~\bibnamefont {Amitay}},\ }\bibfield  {title} {\bibinfo
  {title} {{Coherent Control of Bond Making}},\ }\href
  {https://doi.org/10.1103/PhysRevLett.114.233003} {\bibfield  {journal}
  {\bibinfo  {journal} {Phys. Rev. Lett.}\ }\textbf {\bibinfo {volume} {114}},\
  \bibinfo {pages} {233003} (\bibinfo {year} {2015})}\BibitemShut {NoStop}%
\bibitem [{\citenamefont {Natan}\ \emph {et~al.}(2016)\citenamefont {Natan},
  \citenamefont {Ware}, \citenamefont {Prabhudesai}, \citenamefont {Lev},
  \citenamefont {Bruner}, \citenamefont {Heber},\ and\ \citenamefont
  {Bucksbaum}}]{Light_induced_CI}%
  \BibitemOpen
  \bibfield  {author} {\bibinfo {author} {\bibfnamefont {A.}~\bibnamefont
  {Natan}}, \bibinfo {author} {\bibfnamefont {M.~R.}\ \bibnamefont {Ware}},
  \bibinfo {author} {\bibfnamefont {V.~S.}\ \bibnamefont {Prabhudesai}},
  \bibinfo {author} {\bibfnamefont {U.}~\bibnamefont {Lev}}, \bibinfo {author}
  {\bibfnamefont {B.~D.}\ \bibnamefont {Bruner}}, \bibinfo {author}
  {\bibfnamefont {O.}~\bibnamefont {Heber}},\ and\ \bibinfo {author}
  {\bibfnamefont {P.~H.}\ \bibnamefont {Bucksbaum}},\ }\bibfield  {title}
  {\bibinfo {title} {{Observation of Quantum Interferences via Light-Induced
  Conical Intersections in Diatomic Molecules}},\ }\href
  {https://doi.org/10.1103/PhysRevLett.116.143004} {\bibfield  {journal}
  {\bibinfo  {journal} {Phys. Rev. Lett.}\ }\textbf {\bibinfo {volume} {116}},\
  \bibinfo {pages} {143004} (\bibinfo {year} {2016})}\BibitemShut {NoStop}%
\bibitem [{\citenamefont {Wunderlich}\ \emph {et~al.}(1997)\citenamefont
  {Wunderlich}, \citenamefont {Kobler}, \citenamefont {Figger},\ and\
  \citenamefont {H\"ansch}}]{L_ind_molpots}%
  \BibitemOpen
  \bibfield  {author} {\bibinfo {author} {\bibfnamefont {C.}~\bibnamefont
  {Wunderlich}}, \bibinfo {author} {\bibfnamefont {E.}~\bibnamefont {Kobler}},
  \bibinfo {author} {\bibfnamefont {H.}~\bibnamefont {Figger}},\ and\ \bibinfo
  {author} {\bibfnamefont {T.~W.}\ \bibnamefont {H\"ansch}},\ }\bibfield
  {title} {\bibinfo {title} {{Light-Induced Molecular Potentials}},\ }\href
  {https://doi.org/10.1103/PhysRevLett.78.2333} {\bibfield  {journal} {\bibinfo
   {journal} {Phys. Rev. Lett.}\ }\textbf {\bibinfo {volume} {78}},\ \bibinfo
  {pages} {2333} (\bibinfo {year} {1997})}\BibitemShut {NoStop}%
\bibitem [{\citenamefont {Shaffer}\ \emph {et~al.}(2018)\citenamefont
  {Shaffer}, \citenamefont {Rittenhouse},\ and\ \citenamefont
  {Sadeghpour}}]{Shaffer2018}%
  \BibitemOpen
  \bibfield  {author} {\bibinfo {author} {\bibfnamefont {J.~P.}\ \bibnamefont
  {Shaffer}}, \bibinfo {author} {\bibfnamefont {S.~T.}\ \bibnamefont
  {Rittenhouse}},\ and\ \bibinfo {author} {\bibfnamefont {H.~R.}\ \bibnamefont
  {Sadeghpour}},\ }\bibfield  {title} {\bibinfo {title} {{Ultracold Rydberg
  molecules}},\ }\href {https://doi.org/10.1038/s41467-018-04135-6} {\bibfield
  {journal} {\bibinfo  {journal} {Nat. Commun.}\ }\textbf {\bibinfo {volume}
  {9}},\ \bibinfo {pages} {1965} (\bibinfo {year} {2018})}\BibitemShut
  {NoStop}%
\bibitem [{\citenamefont {Giannakeas}\ \emph {et~al.}(2020)\citenamefont
  {Giannakeas}, \citenamefont {Eiles}, \citenamefont {Robicheaux},\ and\
  \citenamefont {Rost}}]{Dressed_ion_pair_states}%
  \BibitemOpen
  \bibfield  {author} {\bibinfo {author} {\bibfnamefont {P.}~\bibnamefont
  {Giannakeas}}, \bibinfo {author} {\bibfnamefont {M.~T.}\ \bibnamefont
  {Eiles}}, \bibinfo {author} {\bibfnamefont {F.}~\bibnamefont {Robicheaux}},\
  and\ \bibinfo {author} {\bibfnamefont {J.~M.}\ \bibnamefont {Rost}},\
  }\bibfield  {title} {\bibinfo {title} {{Dressed Ion-Pair States of an
  Ultralong-Range Rydberg Molecule}},\ }\href
  {https://doi.org/10.1103/PhysRevLett.125.123401} {\bibfield  {journal}
  {\bibinfo  {journal} {Phys. Rev. Lett.}\ }\textbf {\bibinfo {volume} {125}},\
  \bibinfo {pages} {123401} (\bibinfo {year} {2020})}\BibitemShut {NoStop}%
\bibitem [{\citenamefont {Thomas}\ \emph {et~al.}(2018)\citenamefont {Thomas},
  \citenamefont {Lippe}, \citenamefont {Eichert},\ and\ \citenamefont
  {Ott}}]{Thomas2018}%
  \BibitemOpen
  \bibfield  {author} {\bibinfo {author} {\bibfnamefont {O.}~\bibnamefont
  {Thomas}}, \bibinfo {author} {\bibfnamefont {C.}~\bibnamefont {Lippe}},
  \bibinfo {author} {\bibfnamefont {T.}~\bibnamefont {Eichert}},\ and\ \bibinfo
  {author} {\bibfnamefont {H.}~\bibnamefont {Ott}},\ }\bibfield  {title}
  {\bibinfo {title} {{Experimental realization of a Rydberg optical Feshbach
  resonance in a quantum many-body system}},\ }\href
  {https://doi.org/10.1038/s41467-018-04684-w} {\bibfield  {journal} {\bibinfo
  {journal} {Nat. Commun.}\ }\textbf {\bibinfo {volume} {9}},\ \bibinfo {pages}
  {2238} (\bibinfo {year} {2018})}\BibitemShut {NoStop}%
\bibitem [{\citenamefont {Zou}\ \emph {et~al.}(2023)\citenamefont {Zou},
  \citenamefont {Berngruber}, \citenamefont {Anasuri}, \citenamefont {Zuber},
  \citenamefont {Meinert}, \citenamefont {L\"ow},\ and\ \citenamefont
  {Pfau}}]{atom_ion_dynamics}%
  \BibitemOpen
  \bibfield  {author} {\bibinfo {author} {\bibfnamefont {Y.-Q.}\ \bibnamefont
  {Zou}}, \bibinfo {author} {\bibfnamefont {M.}~\bibnamefont {Berngruber}},
  \bibinfo {author} {\bibfnamefont {V.~S.~V.}\ \bibnamefont {Anasuri}},
  \bibinfo {author} {\bibfnamefont {N.}~\bibnamefont {Zuber}}, \bibinfo
  {author} {\bibfnamefont {F.}~\bibnamefont {Meinert}}, \bibinfo {author}
  {\bibfnamefont {R.}~\bibnamefont {L\"ow}},\ and\ \bibinfo {author}
  {\bibfnamefont {T.}~\bibnamefont {Pfau}},\ }\bibfield  {title} {\bibinfo
  {title} {{Observation of Vibrational Dynamics of Orientated Rydberg-Atom-Ion
  Molecules}},\ }\href {https://doi.org/10.1103/PhysRevLett.130.023002}
  {\bibfield  {journal} {\bibinfo  {journal} {Phys. Rev. Lett.}\ }\textbf
  {\bibinfo {volume} {130}},\ \bibinfo {pages} {023002} (\bibinfo {year}
  {2023})}\BibitemShut {NoStop}%
\bibitem [{\citenamefont {Booth}\ \emph {et~al.}(2015)\citenamefont {Booth},
  \citenamefont {Rittenhouse}, \citenamefont {Yang}, \citenamefont
  {Sadeghpour},\ and\ \citenamefont {Shaffer}}]{booth_lonrange_strong_elDM}%
  \BibitemOpen
  \bibfield  {author} {\bibinfo {author} {\bibfnamefont {D.}~\bibnamefont
  {Booth}}, \bibinfo {author} {\bibfnamefont {S.~T.}\ \bibnamefont
  {Rittenhouse}}, \bibinfo {author} {\bibfnamefont {J.}~\bibnamefont {Yang}},
  \bibinfo {author} {\bibfnamefont {H.~R.}\ \bibnamefont {Sadeghpour}},\ and\
  \bibinfo {author} {\bibfnamefont {J.~P.}\ \bibnamefont {Shaffer}},\
  }\bibfield  {title} {\bibinfo {title} {{Production of trilobite Rydberg
  molecule dimers with kilo-Debye permanent electric dipole moments}},\ }\href
  {https://doi.org/10.1126/science.1260722} {\bibfield  {journal} {\bibinfo
  {journal} {Science}\ }\textbf {\bibinfo {volume} {348}},\ \bibinfo {pages}
  {99} (\bibinfo {year} {2015})}\BibitemShut {NoStop}%
\bibitem [{\citenamefont {Petrosyan}\ and\ \citenamefont
  {M\o{}lmer}(2014)}]{Petrosyan_MW_dressing}%
  \BibitemOpen
  \bibfield  {author} {\bibinfo {author} {\bibfnamefont {D.}~\bibnamefont
  {Petrosyan}}\ and\ \bibinfo {author} {\bibfnamefont {K.}~\bibnamefont
  {M\o{}lmer}},\ }\bibfield  {title} {\bibinfo {title} {{Binding Potentials and
  Interaction Gates between Microwave-Dressed Rydberg Atoms}},\ }\href
  {https://doi.org/10.1103/PhysRevLett.113.123003} {\bibfield  {journal}
  {\bibinfo  {journal} {Phys. Rev. Lett.}\ }\textbf {\bibinfo {volume} {113}},\
  \bibinfo {pages} {123003} (\bibinfo {year} {2014})}\BibitemShut {NoStop}%
\bibitem [{\citenamefont {Peper}\ and\ \citenamefont
  {Deiglmayr}(2020)}]{RF_photodiss_LRM}%
  \BibitemOpen
  \bibfield  {author} {\bibinfo {author} {\bibfnamefont {M.}~\bibnamefont
  {Peper}}\ and\ \bibinfo {author} {\bibfnamefont {J.}~\bibnamefont
  {Deiglmayr}},\ }\bibfield  {title} {\bibinfo {title} {{Photodissociation of
  long-range Rydberg molecules}},\ }\href
  {https://doi.org/10.1103/PhysRevA.102.062819} {\bibfield  {journal} {\bibinfo
   {journal} {Phys. Rev. A}\ }\textbf {\bibinfo {volume} {102}},\ \bibinfo
  {pages} {062819} (\bibinfo {year} {2020})}\BibitemShut {NoStop}%
\bibitem [{\citenamefont {Block}\ and\ \citenamefont
  {Scheel}(2019)}]{Casimir_Macrodiemrs}%
  \BibitemOpen
  \bibfield  {author} {\bibinfo {author} {\bibfnamefont {J.}~\bibnamefont
  {Block}}\ and\ \bibinfo {author} {\bibfnamefont {S.}~\bibnamefont {Scheel}},\
  }\bibfield  {title} {\bibinfo {title} {{Casimir-Polder-induced Rydberg
  macrodimers}},\ }\href {https://doi.org/10.1103/PhysRevA.100.062508}
  {\bibfield  {journal} {\bibinfo  {journal} {Phys. Rev. A}\ }\textbf {\bibinfo
  {volume} {100}},\ \bibinfo {pages} {062508} (\bibinfo {year}
  {2019})}\BibitemShut {NoStop}%
\bibitem [{\citenamefont {Hollerith}\ \emph {et~al.}(2021)\citenamefont
  {Hollerith}, \citenamefont {Rui}, \citenamefont {Rubio-Abadal}, \citenamefont
  {Srakaew}, \citenamefont {Wei}, \citenamefont {Zeiher}, \citenamefont
  {Gross},\ and\ \citenamefont {Bloch}}]{Hollerith_2021}%
  \BibitemOpen
  \bibfield  {author} {\bibinfo {author} {\bibfnamefont {S.}~\bibnamefont
  {Hollerith}}, \bibinfo {author} {\bibfnamefont {J.}~\bibnamefont {Rui}},
  \bibinfo {author} {\bibfnamefont {A.}~\bibnamefont {Rubio-Abadal}}, \bibinfo
  {author} {\bibfnamefont {K.}~\bibnamefont {Srakaew}}, \bibinfo {author}
  {\bibfnamefont {D.}~\bibnamefont {Wei}}, \bibinfo {author} {\bibfnamefont
  {J.}~\bibnamefont {Zeiher}}, \bibinfo {author} {\bibfnamefont
  {C.}~\bibnamefont {Gross}},\ and\ \bibinfo {author} {\bibfnamefont
  {I.}~\bibnamefont {Bloch}},\ }\bibfield  {title} {\bibinfo {title}
  {{Microscopic electronic structure tomography of Rydberg macrodimers}},\
  }\href {https://doi.org/10.1103/PhysRevResearch.3.013252} {\bibfield
  {journal} {\bibinfo  {journal} {Phys. Rev. Research}\ }\textbf {\bibinfo
  {volume} {3}},\ \bibinfo {pages} {013252} (\bibinfo {year}
  {2021})}\BibitemShut {NoStop}%
\bibitem [{\citenamefont {Chin}\ \emph {et~al.}(2010)\citenamefont {Chin},
  \citenamefont {Grimm}, \citenamefont {Julienne},\ and\ \citenamefont
  {Tiesinga}}]{Feshbach_Cheng_2010}%
  \BibitemOpen
  \bibfield  {author} {\bibinfo {author} {\bibfnamefont {C.}~\bibnamefont
  {Chin}}, \bibinfo {author} {\bibfnamefont {R.}~\bibnamefont {Grimm}},
  \bibinfo {author} {\bibfnamefont {P.}~\bibnamefont {Julienne}},\ and\
  \bibinfo {author} {\bibfnamefont {E.}~\bibnamefont {Tiesinga}},\ }\bibfield
  {title} {\bibinfo {title} {{Feshbach resonances in ultracold gases}},\ }\href
  {https://doi.org/10.1103/RevModPhys.82.1225} {\bibfield  {journal} {\bibinfo
  {journal} {Rev. Mod. Phys.}\ }\textbf {\bibinfo {volume} {82}},\ \bibinfo
  {pages} {1225} (\bibinfo {year} {2010})}\BibitemShut {NoStop}%
\bibitem [{\citenamefont {Taylor}(1972)}]{taylor_scattering}%
  \BibitemOpen
  \bibfield  {author} {\bibinfo {author} {\bibfnamefont {J.~R.}\ \bibnamefont
  {Taylor}},\ }\href
  {https://www.bibsonomy.org/bibtex/271e9858a86bc159caa6a0bb03fdec8d1/sknecht}
  {\emph {\bibinfo {title} {{S}cattering {T}heory: {T}he quantum {T}heory on
  {N}onrelativistic {C}ollisions}}}\ (\bibinfo  {publisher} {Wiley, New York},\
  \bibinfo {year} {1972})\BibitemShut {NoStop}%
\bibitem [{SI()}]{SI}%
  \BibitemOpen
  \href@noop {} {\bibinfo {title} {{See Supplementary Material}}}\BibitemShut
  {NoStop}%
\bibitem [{\citenamefont {Weber}\ \emph {et~al.}(2017)\citenamefont {Weber},
  \citenamefont {Tresp}, \citenamefont {Menke}, \citenamefont {Urvoy},
  \citenamefont {Firstenberg}, \citenamefont {B\"uchler},\ and\ \citenamefont
  {Hofferberth}}]{Weber2017}%
  \BibitemOpen
  \bibfield  {author} {\bibinfo {author} {\bibfnamefont {S.}~\bibnamefont
  {Weber}}, \bibinfo {author} {\bibfnamefont {C.}~\bibnamefont {Tresp}},
  \bibinfo {author} {\bibfnamefont {H.}~\bibnamefont {Menke}}, \bibinfo
  {author} {\bibfnamefont {A.}~\bibnamefont {Urvoy}}, \bibinfo {author}
  {\bibfnamefont {O.}~\bibnamefont {Firstenberg}}, \bibinfo {author}
  {\bibfnamefont {H.~P.}\ \bibnamefont {B\"uchler}},\ and\ \bibinfo {author}
  {\bibfnamefont {S.}~\bibnamefont {Hofferberth}},\ }\bibfield  {title}
  {\bibinfo {title} {{Calculation of Rydberg interaction potentials}},\ }\href
  {https://doi.org/10.1088/1361-6455/aa743a} {\bibfield  {journal} {\bibinfo
  {journal} {J. Phys. B: At. Mol. Opt. Phys.}\ }\textbf {\bibinfo {volume}
  {50}},\ \bibinfo {pages} {133001} (\bibinfo {year} {2017})}\BibitemShut
  {NoStop}%
\bibitem [{\citenamefont {Simonelli}\ \emph {et~al.}(2016)\citenamefont
  {Simonelli}, \citenamefont {Valado}, \citenamefont {Masella}, \citenamefont
  {Asteria}, \citenamefont {Arimondo}, \citenamefont {Ciampini},\ and\
  \citenamefont {Morsch}}]{Simonelli_2016}%
  \BibitemOpen
  \bibfield  {author} {\bibinfo {author} {\bibfnamefont {C.}~\bibnamefont
  {Simonelli}}, \bibinfo {author} {\bibfnamefont {M.~M.}\ \bibnamefont
  {Valado}}, \bibinfo {author} {\bibfnamefont {G.}~\bibnamefont {Masella}},
  \bibinfo {author} {\bibfnamefont {L.}~\bibnamefont {Asteria}}, \bibinfo
  {author} {\bibfnamefont {E.}~\bibnamefont {Arimondo}}, \bibinfo {author}
  {\bibfnamefont {D.}~\bibnamefont {Ciampini}},\ and\ \bibinfo {author}
  {\bibfnamefont {O.}~\bibnamefont {Morsch}},\ }\bibfield  {title} {\bibinfo
  {title} {{Seeded excitation avalanches in off-resonantly driven Rydberg
  gases}},\ }\href {https://doi.org/10.1088/0953-4075/49/15/154002} {\bibfield
  {journal} {\bibinfo  {journal} {J. Phys. B: At. Mol. Opt. Phys.}\ }\textbf
  {\bibinfo {volume} {49}},\ \bibinfo {pages} {154002} (\bibinfo {year}
  {2016})}\BibitemShut {NoStop}%
\bibitem [{\citenamefont {Goldschmidt}\ \emph {et~al.}(2016)\citenamefont
  {Goldschmidt}, \citenamefont {Boulier}, \citenamefont {Brown}, \citenamefont
  {Koller}, \citenamefont {Young}, \citenamefont {Gorshkov}, \citenamefont
  {Rolston},\ and\ \citenamefont {Porto}}]{Goldschmidt_2016}%
  \BibitemOpen
  \bibfield  {author} {\bibinfo {author} {\bibfnamefont {E.~A.}\ \bibnamefont
  {Goldschmidt}}, \bibinfo {author} {\bibfnamefont {T.}~\bibnamefont
  {Boulier}}, \bibinfo {author} {\bibfnamefont {R.~C.}\ \bibnamefont {Brown}},
  \bibinfo {author} {\bibfnamefont {S.~B.}\ \bibnamefont {Koller}}, \bibinfo
  {author} {\bibfnamefont {J.~T.}\ \bibnamefont {Young}}, \bibinfo {author}
  {\bibfnamefont {A.~V.}\ \bibnamefont {Gorshkov}}, \bibinfo {author}
  {\bibfnamefont {S.~L.}\ \bibnamefont {Rolston}},\ and\ \bibinfo {author}
  {\bibfnamefont {J.~V.}\ \bibnamefont {Porto}},\ }\bibfield  {title} {\bibinfo
  {title} {{Anomalous Broadening in Driven Dissipative Rydberg Systems}},\
  }\href {https://doi.org/10.1103/PhysRevLett.116.113001} {\bibfield  {journal}
  {\bibinfo  {journal} {Phys. Rev. Lett.}\ }\textbf {\bibinfo {volume} {116}},\
  \bibinfo {pages} {113001} (\bibinfo {year} {2016})}\BibitemShut {NoStop}%
\bibitem [{\citenamefont {de~Léséleuc}\ \emph {et~al.}(2019)\citenamefont
  {de~Léséleuc}, \citenamefont {Lienhard}, \citenamefont {Scholl},
  \citenamefont {Barredo}, \citenamefont {Weber}, \citenamefont {Lang},
  \citenamefont {Büchler}, \citenamefont {Lahaye},\ and\ \citenamefont
  {Browaeys}}]{symmetry_p_phase_2019}%
  \BibitemOpen
  \bibfield  {author} {\bibinfo {author} {\bibfnamefont {S.}~\bibnamefont
  {de~Léséleuc}}, \bibinfo {author} {\bibfnamefont {V.}~\bibnamefont
  {Lienhard}}, \bibinfo {author} {\bibfnamefont {P.}~\bibnamefont {Scholl}},
  \bibinfo {author} {\bibfnamefont {D.}~\bibnamefont {Barredo}}, \bibinfo
  {author} {\bibfnamefont {S.}~\bibnamefont {Weber}}, \bibinfo {author}
  {\bibfnamefont {N.}~\bibnamefont {Lang}}, \bibinfo {author} {\bibfnamefont
  {H.~P.}\ \bibnamefont {Büchler}}, \bibinfo {author} {\bibfnamefont
  {T.}~\bibnamefont {Lahaye}},\ and\ \bibinfo {author} {\bibfnamefont
  {A.}~\bibnamefont {Browaeys}},\ }\bibfield  {title} {\bibinfo {title}
  {{Observation of a symmetry-protected topological phase of interacting bosons
  with Rydberg atoms}},\ }\href {https://doi.org/10.1126/science.aav9105}
  {\bibfield  {journal} {\bibinfo  {journal} {Science}\ }\textbf {\bibinfo
  {volume} {365}},\ \bibinfo {pages} {775} (\bibinfo {year}
  {2019})}\BibitemShut {NoStop}%
\bibitem [{\citenamefont {Chew}\ \emph {et~al.}(2022)\citenamefont {Chew},
  \citenamefont {Tomita}, \citenamefont {Mahesh}, \citenamefont {Sugawa},
  \citenamefont {de~L{\'e}s{\'e}leuc},\ and\ \citenamefont
  {Ohmori}}]{Chew2022}%
  \BibitemOpen
  \bibfield  {author} {\bibinfo {author} {\bibfnamefont {Y.}~\bibnamefont
  {Chew}}, \bibinfo {author} {\bibfnamefont {T.}~\bibnamefont {Tomita}},
  \bibinfo {author} {\bibfnamefont {T.~P.}\ \bibnamefont {Mahesh}}, \bibinfo
  {author} {\bibfnamefont {S.}~\bibnamefont {Sugawa}}, \bibinfo {author}
  {\bibfnamefont {S.}~\bibnamefont {de~L{\'e}s{\'e}leuc}},\ and\ \bibinfo
  {author} {\bibfnamefont {K.}~\bibnamefont {Ohmori}},\ }\bibfield  {title}
  {\bibinfo {title} {{Ultrafast energy exchange between two single Rydberg
  atoms on a nanosecond timescale}},\ }\href
  {https://doi.org/10.1038/s41566-022-01047-2} {\bibfield  {journal} {\bibinfo
  {journal} {Nat. Photonics}\ }\textbf {\bibinfo {volume} {16}},\ \bibinfo
  {pages} {724} (\bibinfo {year} {2022})}\BibitemShut {NoStop}%
\bibitem [{\citenamefont {Marcuzzi}\ \emph {et~al.}(2017)\citenamefont
  {Marcuzzi}, \citenamefont {Min\'a\ifmmode~\check{r}\else \v{r}\fi{}},
  \citenamefont {Barredo}, \citenamefont {de~L\'es\'eleuc}, \citenamefont
  {Labuhn}, \citenamefont {Lahaye}, \citenamefont {Browaeys}, \citenamefont
  {Levi},\ and\ \citenamefont {Lesanovsky}}]{Facilitation_Antoine}%
  \BibitemOpen
  \bibfield  {author} {\bibinfo {author} {\bibfnamefont {M.}~\bibnamefont
  {Marcuzzi}}, \bibinfo {author} {\bibfnamefont {J.~c.~v.}\ \bibnamefont
  {Min\'a\ifmmode~\check{r}\else \v{r}\fi{}}}, \bibinfo {author} {\bibfnamefont
  {D.}~\bibnamefont {Barredo}}, \bibinfo {author} {\bibfnamefont
  {S.}~\bibnamefont {de~L\'es\'eleuc}}, \bibinfo {author} {\bibfnamefont
  {H.}~\bibnamefont {Labuhn}}, \bibinfo {author} {\bibfnamefont
  {T.}~\bibnamefont {Lahaye}}, \bibinfo {author} {\bibfnamefont
  {A.}~\bibnamefont {Browaeys}}, \bibinfo {author} {\bibfnamefont
  {E.}~\bibnamefont {Levi}},\ and\ \bibinfo {author} {\bibfnamefont
  {I.}~\bibnamefont {Lesanovsky}},\ }\bibfield  {title} {\bibinfo {title}
  {{Facilitation Dynamics and Localization Phenomena in Rydberg Lattice Gases
  with Position Disorder}},\ }\href
  {https://doi.org/10.1103/PhysRevLett.118.063606} {\bibfield  {journal}
  {\bibinfo  {journal} {Phys. Rev. Lett.}\ }\textbf {\bibinfo {volume} {118}},\
  \bibinfo {pages} {063606} (\bibinfo {year} {2017})}\BibitemShut {NoStop}%
\bibitem [{\citenamefont {Kim}\ \emph {et~al.}(2023)\citenamefont {Kim},
  \citenamefont {Yang}, \citenamefont {Mølmer},\ and\ \citenamefont
  {Ahn}}]{kim2023realization}%
  \BibitemOpen
  \bibfield  {author} {\bibinfo {author} {\bibfnamefont {K.}~\bibnamefont
  {Kim}}, \bibinfo {author} {\bibfnamefont {F.}~\bibnamefont {Yang}}, \bibinfo
  {author} {\bibfnamefont {K.}~\bibnamefont {Mølmer}},\ and\ \bibinfo {author}
  {\bibfnamefont {J.}~\bibnamefont {Ahn}},\ }\href@noop {} {\bibinfo {title}
  {{Realization of an extremely anisotropic Heisenberg magnet in Rydberg atom
  arrays}}} (\bibinfo {year} {2023}),\ \Eprint
  {https://arxiv.org/abs/2307.04342} {arXiv:2307.04342} \BibitemShut {NoStop}%
\bibitem [{\citenamefont {Mazza}\ \emph {et~al.}(2020)\citenamefont {Mazza},
  \citenamefont {Schmidt},\ and\ \citenamefont
  {Lesanovsky}}]{kin_const_facilit}%
  \BibitemOpen
  \bibfield  {author} {\bibinfo {author} {\bibfnamefont {P.~P.}\ \bibnamefont
  {Mazza}}, \bibinfo {author} {\bibfnamefont {R.}~\bibnamefont {Schmidt}},\
  and\ \bibinfo {author} {\bibfnamefont {I.}~\bibnamefont {Lesanovsky}},\
  }\bibfield  {title} {\bibinfo {title} {{Vibrational Dressing in Kinetically
  Constrained Rydberg Spin Systems}},\ }\href
  {https://doi.org/10.1103/PhysRevLett.125.033602} {\bibfield  {journal}
  {\bibinfo  {journal} {Phys. Rev. Lett.}\ }\textbf {\bibinfo {volume} {125}},\
  \bibinfo {pages} {033602} (\bibinfo {year} {2020})}\BibitemShut {NoStop}%
\bibitem [{\citenamefont {Magoni}\ \emph {et~al.}(2023)\citenamefont {Magoni},
  \citenamefont {Joshi},\ and\ \citenamefont
  {Lesanovsky}}]{Matteo_jahn_teller}%
  \BibitemOpen
  \bibfield  {author} {\bibinfo {author} {\bibfnamefont {M.}~\bibnamefont
  {Magoni}}, \bibinfo {author} {\bibfnamefont {R.}~\bibnamefont {Joshi}},\ and\
  \bibinfo {author} {\bibfnamefont {I.}~\bibnamefont {Lesanovsky}},\
  }\href@noop {} {\bibinfo {title} {{Rydberg tweezer molecules: Spin-phonon
  entanglement and Jahn-Teller effect}}} (\bibinfo {year} {2023}),\ \Eprint
  {https://arxiv.org/abs/2303.08861} {arXiv:2303.08861} \BibitemShut {NoStop}%
\bibitem [{\citenamefont {{Sherson}}\ \emph {et~al.}(2010)\citenamefont
  {{Sherson}}, \citenamefont {{Weitenberg}}, \citenamefont {{Endres}},
  \citenamefont {{Cheneau}}, \citenamefont {{Bloch}},\ and\ \citenamefont
  {{Kuhr}}}]{Sherson2010}%
  \BibitemOpen
  \bibfield  {author} {\bibinfo {author} {\bibfnamefont {J.~F.}\ \bibnamefont
  {{Sherson}}}, \bibinfo {author} {\bibfnamefont {C.}~\bibnamefont
  {{Weitenberg}}}, \bibinfo {author} {\bibfnamefont {M.}~\bibnamefont
  {{Endres}}}, \bibinfo {author} {\bibfnamefont {M.}~\bibnamefont {{Cheneau}}},
  \bibinfo {author} {\bibfnamefont {I.}~\bibnamefont {{Bloch}}},\ and\ \bibinfo
  {author} {\bibfnamefont {S.}~\bibnamefont {{Kuhr}}},\ }\bibfield  {title}
  {\bibinfo {title} {{Single-atom-resolved fluorescence imaging of an atomic
  Mott insulator}},\ }\href {https://doi.org/10.1038/nature09378} {\bibfield
  {journal} {\bibinfo  {journal} {Nature}\ }\textbf {\bibinfo {volume} {467}},\
  \bibinfo {pages} {68} (\bibinfo {year} {2010})}\BibitemShut {NoStop}%
\bibitem [{\citenamefont {\u{S}ibali\'c}\ \emph {et~al.}(2017)\citenamefont
  {\u{S}ibali\'c}, \citenamefont {Pritchard}, \citenamefont {Adams},\ and\
  \citenamefont {Weatherill}}]{Sibalic2017}%
  \BibitemOpen
  \bibfield  {author} {\bibinfo {author} {\bibfnamefont {N.}~\bibnamefont
  {\u{S}ibali\'c}}, \bibinfo {author} {\bibfnamefont {J.~D.}\ \bibnamefont
  {Pritchard}}, \bibinfo {author} {\bibfnamefont {C.~S.}\ \bibnamefont
  {Adams}},\ and\ \bibinfo {author} {\bibfnamefont {K.~J.}\ \bibnamefont
  {Weatherill}},\ }\bibfield  {title} {\bibinfo {title} {{ARC: An open-source
  library for calculating properties of alkali Rydberg atoms}},\ }\href
  {https://doi.org/10.1016/j.cpc.2017.06.015} {\bibfield  {journal} {\bibinfo
  {journal} {Computer Physics Communications}\ }\textbf {\bibinfo {volume}
  {220}},\ \bibinfo {pages} {319} (\bibinfo {year} {2017})}\BibitemShut
  {NoStop}%
\end{thebibliography}%

\section*{Supplementary material}
This supplementary material discusses further theoretical and experimental details and contains additional experimental datasets supporting the claims made in the main text.

\subsection{Theory}

\subsubsection{Macrodimers on a lattice}
To describe the combined effects of macrodimer binding and light coupling, we use the following model: We consider $N$ atoms distributed in one dimension at positions $x_i$, excited by the probe and the control laser. Assuming that the probe laser is weak, we restrict our attention to the space of states reached by a single probe photon: These comprise states with a single Rydberg excitation as well as states with a single macrodimer excitation. A singly-excited state with a Rydberg excitation on site $i$ is described by 
\begin{align}\label{eq:singly_ex_st}
 |\widetilde{\Phi}_i\rangle = \rho_i (\mathcal{R},
 R_1,...,R_{N-1}) |e_i\rangle.
\end{align}
Here, we introduced the total motional state $\rho_i (\mathcal{R},
 R_1,...,R_{N-1})$ written in position space, as denoted by the tilde. The motional state is parametrized in a convenient set of ``links'', i.e. of relative coordinates $\vec{R} = (R_1, R_2,...,R_{N-1})$ with $R_\lambda = x_\lambda - x_{\lambda+1}$ (here, the subindex $\lambda$ denotes the link between sites $\lambda$ and $\lambda+1$) and the center of mass coordinate, $\mathcal{R} = 1/N \sum_i x_i$, as well as the notation $|e_i\rangle \equiv |g_1 ... e_i...g_{N}\rangle$ for the electronic state. The macrodimer state with its electronic wave function $\ket{\Psi_{\mathrm{el}, \,\lambda}^{(2)}}$ has a fixed motional relative wave function $ \Phi_{v}(R_\lambda)$, the index $v$ labels the vibrational macrodimer mode. 
We consider many-body states containing a macrodimer excitation that form a product state with the state of the remaining ground state atoms in the motional state $\varphi(\mathcal{R}, R_1, ..., R_{\lambda-1}, R_{\lambda+1},...,R_{N-1})$, yielding
\begin{align}\label{eq:single_md_st}
 |\widetilde{\Psi}^\lambda_v \rangle =\, & \Phi_{v}(R_\lambda)\ket{\Psi_{\mathrm{el}, \,\lambda}^{(2)}} \varphi(\mathcal{R}, R_1, ..., R_{\lambda-1}, R_{\lambda+1},..., R_{N-1}) \notag \\  & \times |g_1...g_{\lambda-1} g_{\lambda+2} ...g_2\rangle.
\end{align}
The kinetic energy of the atoms is
\begin{align}\label{eq:Ekin}
 \mathcal{T} &= -\frac{\hbar^2}{2m} \sum_i \frac{d^2}{dx_i^2} \\
 &= -\frac{\hbar^2}{2m} \left( \frac{1}{N} \frac{d^2}{d\mathcal{R}^2} + 2 \sum_{\lambda=1}^{N-1} \frac{d^2}{dR_\lambda^2} - 2 \sum_{\lambda=2}^{N-1} \frac{d}{dR_\lambda} \frac{d}{dR_{\lambda-1}} \right).\notag
\end{align}
Neglecting the effects of the optical lattice, all atoms that are not bound in macrodimer states can freely disperse. We assume that the macrodimer states are well approximated by harmonic oscillator eigenstates. 
It is convenient to expand all states in a basis of momentum eigenstates (denoted without a tilde), with $K$ representing the center of mass and $k_\lambda$ the relative coordinate $R_\lambda$. We use the shorthand $\vec{k}_{\perp \lambda}=(k_1,...,k_{\lambda-1},k_{\lambda+1},...,k_{N-1})$ to denote the set of relative momenta different from $k_\lambda$. 

The total Hamiltonian $\mathcal{H} =\mathcal{T} + \mathcal{V} + \mathcal{H}_L$ contains the kinetic energy, the macrodimer binding potential $\mathcal{V}$, as well as the term $\mathcal{H}_L$ which couples the singly-excited states to the macrodimer states by laser light.
Different singly-excited states are eigenstates of the atomic Hamiltonian and not coupled by the laser. 
Thus, within their subspace, the Hamiltonian is diagonal and only contributes via kinetic terms, with matrix elements
\begin{equation}
\begin{aligned}
 &\langle \Phi_i (K,\vec{k}) | \mathcal{H} | \Phi_{i'} (K',\vec{k}') \rangle = \\ &\delta_{i,i'} \delta_{K,K'} \delta_{\vec{k},\vec{k}'} \frac{\hbar^2}{2m} \left( 2 \sum_{\lambda=1}^{N-1} k_\lambda^2 - 2 \sum_{\lambda=2}^{N-1} k_\lambda k_{\lambda-1} + \frac{K^2}{N} \right)
\end{aligned}
\end{equation}
in the frame co-rotating with the coupling laser. The situation is slightly more complex for states involving a macrodimer: Here, the mode coupling of the kinetic energies in the relative coordinates (see Eq.~(\ref{eq:Ekin})) mixes macrodimer states for $N\geq3$. However, this direct motional coupling is small and we restrict our basis for the consideration of the case of  three atoms ($N=3$) to a single macrodimer state, such that the last term of the following equation vanishes. 
\begin{widetext}
\begin{equation}\label{eq:energy_macrodimer}
\begin{aligned} 
 \langle \Psi^{\bar{\lambda}}_v (K,\vec{k}_{\perp \bar{\lambda}}) | \mathcal{H} | \Psi^{\bar{\lambda}'}_{v'} (K',\vec{k}'_{\perp \bar{\lambda}'}) \rangle = 
 \delta_{\bar{\lambda}, \bar{\lambda}'} \delta_{v, v'} \delta_{K,K'} \delta_{\vec{k}_{\perp \lambda},\vec{k}'_{\perp \lambda}} \left( \Delta^v_\mathrm{C} +\frac{\hbar^2}{2m} \left( 2 \sum_{\substack{\lambda=1 \\ \lambda \neq \bar\lambda}}^{N-1} k_\lambda^2  + \frac{K^2}{N}\right) \right) \\- 2 \frac{\hbar^2}{2m}\sum_{\substack{\lambda=2 \\ \lambda \not\in \{\bar{\lambda}, \bar{\lambda}+1\}}}^{N-1}  \langle \Psi^{\bar{\lambda}}_v (K,\vec{k}_{\perp \bar{\lambda}}) |k_\lambda k_{\lambda-1} | \Psi^{\bar{\lambda}'}_{v'} (K',\vec{k}'_{\perp \bar{\lambda}'}) \rangle.
\end{aligned}
\end{equation}
\end{widetext}
We chose the energy of the intermediate states at rest as $E=0$ and defined the macrodimer detuning $\Delta^v_\mathrm{C}$ with respect to the macrodimer energy (see section~\ref{section:exc_scheme}). The macrodimer detuning $\Delta^v_\mathrm{C}$ reflects the relative kinetic energy as well as the macrodimer binding potential on the respective link. In contrast to the rest of the manuscript (see section~\ref{section:exc_scheme}), we introduced another superindex $v$ in order to also include higher vibrational macrodimer modes with $v \neq 0$.

Finally, we evaluate the optical coupling between singly-excited states Eq.~\ref{eq:singly_ex_st} and macrodimer states Eq.~\ref{eq:single_md_st} in momentum space. The laser only changes the electronic state but transfers no significant momentum, making the center-of-mass kinetic energy a constant of motion. 
The corresponding matrix elements are given by
\begin{equation}
\begin{aligned}
  &\langle \Phi_i (K,\vec{k}) | \mathcal{H} |  \Psi^{\lambda}_v (K,\vec{k}_{\perp \lambda})  \rangle = \\
 &\langle e_i | \mathcal{H}_L \ket{\Psi_{\mathrm{el}, \,\lambda}^{(2)}} \int \rho_i^*(\vec{k}) \varphi(\vec{k}_{\perp k_{\lambda}} ) \phi_v(k_{\lambda})  \varrho_k  \ \rm d \vec{k}
\end{aligned}
\end{equation}
with the density of states $\varrho_k$. The first term in the second line only contributes when $\lambda = i$ or $\lambda = i+1$. We identify $\langle e_i | \mathcal{H}_L  \ket{\Psi_{\mathrm{el}, \,\lambda}^{(2)}} = \frac{1}{2} \alpha \hbar\mathrm\Omega_{ge}$, see section~\ref{section:scheme_2} and section~\ref{section_pert}. The second term is the Franck-Condon overlap. 

\subsubsection{Eigenstates, Phases and Spectra}
\begin{figure}
  \centering
  \includegraphics[width=1.0\columnwidth]{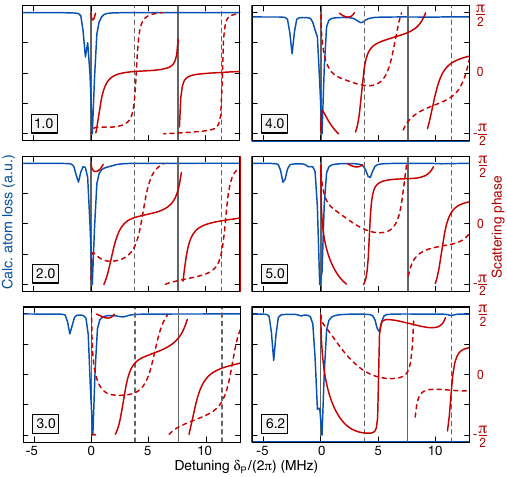}%
  \caption{\label{fig:Fano_phase_theory}\textbf{Calculated absorption profile and scattering phase evolution.} The coupling rate $\mathrm\Omega_C/2\pi$ increases from top to bottom as well as left to right, as indicated in the lower left of the plots. Across positive energies, the evolution of the scattering phase of the continuum states with even (odd) symmetry is shown in solid (dashed) red. At vanishingly small $\mathrm\Omega_C$, we find regions of rapid phase evolution mainly at the energies of the original even (odd) vibrational states, indicated in solid (dashed) gray lines. At intermediate couplings, the phase evolution broadens, indicating accelerated decay of the macrodimer states. At large couplings, regions of rapid phase evolution shift to the energies of the emerging macrodimeron states, indicating that the positive macrodimeron becomes quasi-stable. This sharpening is also also reflected in the atom loss spectra, which are dominated by the even eigenstates. In contrast to Fig.~\ref{fig:2}\,\textbf{(b)} in the main text, where only a single vibrational mode has been included, the two-atom model implemented here included the six lowest vibrational modes, see also Fig.~\ref{fig:SI_S3}. The broadening of the central line towards negative detunings for larger $\mathrm\Omega_C/2\pi$ can also be observed in the experimental datasets (see Fig.~\ref{fig:2}\,\textbf{(a)} and Fig.~\ref{fig:SI_S5}\,\textbf{(a)}) and originates from negative macrodimerons of higher vibrational modes.} 
  \end{figure}
The above model is solved by exact diagonalization on a finite grid in momentum space, yielding the spectrum of eigenenergies $E$ and corresponding eigenstates $|\Psi(E)\rangle$. The eigenstates carry contributions of the macrodimer's electronic states as well as the intermediate states, both with motional wave functions. 
The negative macrodimerons are shifted out of the continuum, which is energetically bounded from below, and therefore do not experience any resonant coupling to free continuum states. For positive energies, each eigenstate is associated with a scattering phase, i.e. the phase shift of the continuum modes induced by the resonant coupling to the macrodimer potential, at distances far away from the potential.
We compute this phase by transforming the states $|\Psi(E)\rangle$ to real space, where the phase can be extracted from the asymptotic wave function $\cos(kr) + K\sin(kr)$ and $\delta = -\arctan(K)$, with $k = \sqrt{2\frac{m E}{2\hbar^2}}$. Since we employ a parabolic approximation of the macrodimer binding potential, all eigenstates are either of even or odd parity with respect to the position of the minimum of the macrodimer binding potential. In Fig.~\ref{fig:Fano_phase_theory}, we illustrate the phases for a system of two atoms 
at varying coupling rates $\mathrm\Omega_C$. At low $\mathrm\Omega_C$, even-parity scattering states couple only to the even macrodimer states, showing rapid phase evolution around each of the even macrodimer states. Likewise, the phase of the odd-parity states exhibit a similiar $\pi$ phase shift at the energies of the odd-parity macrodimer states. We note that this behavior is fully equivalent to Feshbach resonances, where isolated macrodimer states can be viewed as bound states of the closed channel. Since the lattice ground state is nearly constant over the length scale of the vibrational wave functions and much more extended than the modulations of the scattering states, the odd-parity states do not contribute significantly to the optical signal (discussed below).

Switching the coupling field $\mathrm\Omega_C$ on first widens the narrow macrodimer resonances, see the scattering phase evolution in Fig.~\ref{fig:Fano_phase_theory}. This can interpreted as a signature of accelerated decay of the bound macrodimer state into the continuum or, equivalently, a reduced lifetime of the macrodimer~\cite{taylor_scattering}. This is accompanied with a shift of the macrodimeron resonance position, in full accordance with experimental observations. As the optical coupling strength is increased even further, we observe that the phase evolution sharpens around the positive-macrodimeron. This indicates that the macodimeron is becoming a quasi-bound state with an increased lifetime. 


The optical spectra are computed from overlaps of the eigenstates with the lattice ground state.
Since both the light coupling and the vibrational macrodimer energies are much larger than the kinetic energy in the optical lattice, the lattice has been neglected in the Hamiltonian. 
However, when probing the obtained eigenstates $|\Psi(E)\rangle$ from the lattice ground states, the optical lattice potential defines the initial motional states of the atoms, which contribute to the expected loss spectra via a Franck-Condon integral. 
The initial motional states of the individual atoms in the trap can be well approximated by a Gaussian of a spatial width $\sigma = \hbar/(2m\omega_{\mathrm{lat}})^{1/2}$ (see also section~\ref{section:scales}) centered at $x_i = i \cdot \sqrt{2}a_{\mathrm{lat}}$, with the diagonal lattice separation $\sqrt{2}a_{\mathrm{lat}}$. The global ground state $| \widetilde{\Phi}_g(x_1,...,x_N) \rangle $ is a product state of the states of the individual atoms and can be factorized in momentum space into 
\begin{align}
 |\Phi_g(K,\vec{k})\rangle = \Phi_{\mathrm{cm}}(K) \Phi_{\mathrm{rel}}(\vec{k}) |g_1...g_N\rangle
\end{align}
with the center-of-mass and relative wave functions
\begin{align}
 \Phi_{\mathrm{cm}}(K) &= \left( \frac{2 \sigma^2}{N \pi} \right)^\frac{1}{4}  e^{ia \frac{N+1}{2}K - \frac{\sigma^2}{N}K^2} \\
 \Phi_{\mathrm{rel}}(\vec{k}) &= N^\frac{1}{4}\left( \frac{2 \sigma^2}{\pi} \right)^\frac{N-1}{4}e^{- 2\sigma^2 \sum_{i=1}^{N-1}k_i^2} \\ &\ \ \ \ \cdot e^{ 2\sigma^2\sum_{i=2}^{N-1} k_ik_{i-1} } e^{ia\sum_{j=1}^{N-1} k_j} \nonumber.
\end{align}
The probe field absorption is now directly given by overlaps with the probe Hamiltonian $\mathcal{H}_P$
\begin{align} \label{eq:abs_spectrum}
 C_\text{abs}(E) \propto | \langle \Psi(E) | \mathcal{H}_P | \Phi_g(K,\vec{k}) \rangle |^2.
\end{align}
The optically active part of the eigenstates are the singly-excited states since only they are optically coupled from the ground state. Since the center-of-mass motion is factorizable and the initial state is in a product state, the center-of-mass wave function will not get entangled with the relative motion.

In the experiments, the observable is the loss rate from the trap after a variable delay time. Since the exact dynamics of this process are complex, depending on the trap depth, the anti-trapping effect of the optical lattice on the Rydberg states, and the spread into three-dimensional rotational states of the macrodimer, we assume that none of the atoms excited to a Rydberg state but all of those in the electronic ground state will be recaptured. This yields a variable loss fraction $f(E)$ for each eigenstate and an overall atom loss signal $S(E) \propto f(E) C_\text{abs}(E)$. 
Fig.~\ref{fig:2}\,\textbf{(b)} and Fig.~\ref{fig:3}\,\textbf{(b)} in the main text are obtained from the signal after smoothening the function with a Gaussian of width of $150\,\si{\kilo\hertz}$ to estimate the additional broadening from the laser. In the same figures, we also illustrate the motional wavefunction in the electronic sector of the singly-excited states. For each of the resonances, these are coherently averaged over a small energy window to represent the optically excited motional states.

Note that assuming only Rydberg excitations to be ejected from the lattice underestimates the signal strengths at the macrodimeron and macrotrimeron resonances due to the illustrated quenched motional wave functions. Under this assumption, the calculated absorption profile neglects the additional loss of ground state atoms in quenched motional states, as would be expected at dimeron and trimeron resonances. Even at the central resonance where only singly-excited states contribute, the modified motional states may occasionally induce multi-atom loss. 

\subsubsection{Perturbations in the binding potential}\label{section_pert}
\begin{figure}
  \centering
  \includegraphics[width=1.0\columnwidth]{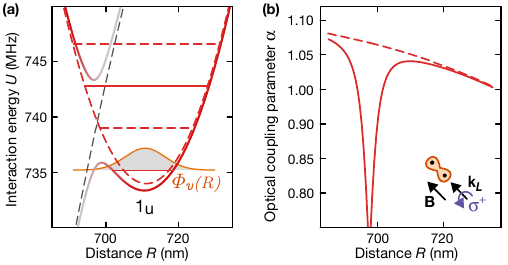}
  \caption{\label{fig_pot_pert}\textbf{Details of the binding potential.} \textbf{(a)} The optically coupled diabatic binding potential (dashed red potential) is crossed by another diabatic $1_u$ potential which is not coupled by the light (dashed gray potential). Dipole-quadrupole interactions between them induce a gap and provide two new adiabatic pair potentials (solid potential curves). Solid (dashed) vertical lines indicate the energies of even (odd) vibrational modes in the diabatic binding potential. We mainly focused on the lowest vibrational mode $\Phi_v(R)$. \textbf{(b)} Due to $R$-dependent electronic structure of the two potentials, the optical coupling parameter $\alpha (R)$ depends on $R$. 
  The dependence is small for the diabatic potential (dashed red line) where $\alpha \approx 1.04$ but significant for the two adiabatic potentials (two solid lines).
  The relative orientation between the interatomic axis, the magnetic field $\mathbf{B}$, and the light polarization $\mathbfit{\varepsilon}$ were chosen as specified in Sec.~\ref{Sec:SI_details}.}
\end{figure}
In addition to higher vibrational modes, two other effects may explain the remaining deviations between the calculated and the observed resonance position in Fig.~\ref{fig:2}\,\textbf{(a)}.

First, a second optically uncoupled pair potential crosses our binding potential close to the binding potential minimum, see Fig.~\ref{fig_pot_pert}\,\textbf{(a)}. 
The finite coupling between both potentials induces a gap at the crossing point energetically similar to the vibrational energy in the binding potential.
Here, the Born-Oppenheimer approximation breaks down and motional states hosted by both crossing potentials are mixed. 
While previous spectra of the same $1_u$ binding potential showed that the vibrational motion follows the gap mostly diabatically, a broadening of the second vibrational level close to the gap indicates that the effect cannot be fully neglected, see Fig.~3\,\textbf{(a)} in Ref.~\cite{Hollerith_2021}. 

Second, the electronic structure of the binding potential is not constant over the length scale of the vibrational mode $\Phi_v(R)$, resulting in a Rabi coupling $\mathrm\Omega_\mathrm{C}(R) = \alpha (R)\mathrm\Omega_{ge}$ that depends on the interatomic distance $R$.
As a consequence, the laser can mix the vibrational eigenmodes of the binding potential into new motional modes at shifted energies.
The coupling coefficient $\alpha (R)$ can be calculated using the $R$-dependent expansion coefficients $c_{ij}(R)$ of the electronic macrodimer state $\ket{\Psi_{\mathrm{el}}^{(2)}} = \sum_{ij} c_{ij}(R) \ket{r_ir_j}$ into non-interacting Rydberg pair states $\ket{r_ir_j}$, see Fig.~\ref{fig_pot_pert}\,\textbf{(b)}.

\subsection{Experimental details}\label{Sec:SI_details}
All experiments started with $^{87}$Rb atoms in the electronic hyperfine ground state $\ket{g} = \ket{F=2,m_F=-2}$, arranged in a two-dimensional optical square lattice with lattice constant $a_{\mathrm{lat}}=532\,\si{\nano\meter}$ at initial filling of $ 90(5)\,\%$ and prepared in the motional ground state of the on-site traps using the Mott insulating phase.
The propagation direction of the $\sigma^+$-polarized Rydberg excitation laser at an ultraviolet (UV) wavelength $\lambda = 298\,\si{\nano\meter}$ was parallel to the chosen bias field $|\mathbf{B}| = 0.5\,\mathrm{G}$.

\subsubsection{Excitation scheme}\label{section:exc_scheme}
In our excitation scheme, the probe field at a laser frequency $\nu_{P}$ was provided by the red sideband of a phase-modulated UV laser.
It had a small detuning $\delta_\mathrm{p}/2\pi = \nu_{P} - \nu_{g}^{e}$ from the UV transition $\ket{g}\rightarrow \ket{e} = \ket{36P_{1/2},m_J=+1/2}$ at a transition frequency $ \nu_{g}^{e}$.
The large modulation frequency was chosen to be close to the interaction shift $U=735.2\,\si{\mega\hertz}$ of the $1_u$ binding potential from the energy of the non-interacting state $\ket{ee}$\,\cite{Hollerith_2021}.
The carrier field had a large detuning $\Delta^{ge}_\mathrm{C}/2\pi = \nu_{C} - \nu_{g}^{e} \approx U$ from the $\ket{g} \rightarrow \ket{e}$ transition.
Furthermore, the carrier frequency $\nu_C \approx \nu_P + 735\,\si{\mega\hertz}$ was close to the transition frequency $\nu_{ge}^{\Psi_v} = \nu_{g}^{e} + U$ from singly Rydberg-excited states into the doubly-excited macrodimer state $\ket{ge},\ket{eg} \rightarrow \ket{\Psi_v}$.
The remaining detuning $\mathrm\Delta_\mathrm{C}/2\pi = \nu_C - \nu_{ge}^{\Psi_v}$ as well as $\delta_\mathrm{p}$ were tunable by the frequency of the carrier and the sideband\,\cite{Hollerith_2021}.
The blue sideband did not contribute to the experiments. 

\subsubsection{Details on the macrodimer state}\label{section:scheme_2}
The calculated bond length $R_v = 712(5)\,\si{\nano\meter} \approx \sqrt{2}a_{\mathrm{lat}}$ was close to the lattice diagonal distance and macrodimers could be excited at distances $\mathbf{R}_0=(-1,1)\,a_{\mathrm{lat}}$ as well as $\mathbf{R}_\perp=(-1,-1)\,a_{\mathrm{lat}}$. 
The excited $1_u$ macrodimers had a total angular momentum projection $M = \pm 1$ of both atoms along the interatomic axis. 
For orientations $\mathbf{R}_\perp$, macrodimer excitation rates for $M=\pm 1$ were finite but strongly supressed and therefore neglected~\cite{Hollerith_2021}.
Along the diagonal direction $\mathbf{R}_0 \parallel \mathbf{B}$, dipole selection rules only allow for the excitation of the macrodimer state with projection $M = +1$. 
For this orientation, we calculate the carrier Rabi frequency of the electronic coupling as $\mathrm\Omega_\mathrm{C} = \alpha\mathrm\Omega_{ge}$, with $\alpha \approx 1.04$ accounting for the electronic structure of the macrodimer state and $\mathrm\Omega_{ge}$ the experimentally calibrated single-atom Rabi frequency between $\ket{g}$ and $\ket{e}$~\cite{Hollerith_2021}. 
The probe Rabi coupling $\mathrm\Omega_\mathrm{p}$ coupling $\ket{g}$ and $\ket{e}$ was smaller than $\mathrm\Omega_\mathrm{C}$ ($\mathrm\Omega_\mathrm{p}\leq 1/10 \, \mathrm\Omega_\mathrm{C}$ for the observed spectra and the correlations observed in Fig.~\ref{fig:3} and $\mathrm\Omega_\mathrm{p}\leq 1/3 \, \mathrm\Omega_\mathrm{C}$ for all other correlation measurements).

\subsubsection{Energy scales and dissipation}\label{section:scales}
Here, we discuss the energy scales and dissipation mechanisms present in our system.
The dominant energy scales were $\mathrm\Omega_\mathrm{C}$ and the kinetic energy of the lowest vibrational macrodimer mode $E_{\mathrm{kin}}/h =  850\,\si{\kilo\hertz}$, which is one quarter of the vibrational frequency $\omega_v/2\pi = 3.8\,\si{\mega\hertz}$ in the binding potential. 
Atom pairs released from the lowest vibrational mode into free space move with a high relative velocity $\delta v \approx \sqrt{h \omega_v/m} = 125\,\si{\nano \meter}/\si{\micro \second}$ along their relative coordinate $R$, with $h$ the Planck constant and $m$ the mass of $^{87}$Rb.
At the chosen lattice depth of $1000\,E_r$ along all three directions, the on-site oscillation frequency $\omega_{\mathrm{lat}}/2\pi = 128\,\si{\kilo\hertz}$, was significantly smaller than $\omega_v$.
Here, $E_r = h^2/(8ma_{\mathrm{lat}}^2)$ is the recoil energy of the optical lattice~\cite{Sherson2010}.
The initial ground state relative wave function $\Phi_g(R)$ is about eight times broader than the mode $\Phi_v(R)$. 
Because macrodimer excitation only compresses the relative coordinates and keeps other coordinates unaffected, we neglected motion perpendicular to the direction of macrodimer excitation and treated the system as one dimensional, as presented in in the main text.
Furthermore, we neglected the center of mass coordinate.
 \begin{figure*}
  \centering
  \includegraphics[width=1.0\textwidth]{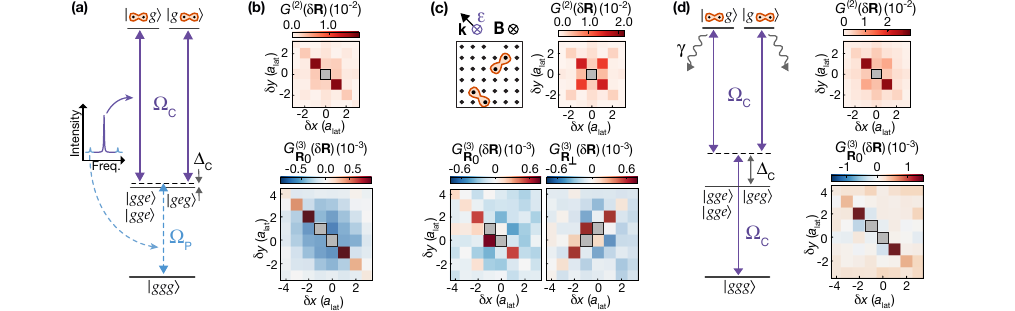}%
  \caption{\label{fig:S1}\textbf{Further microscopic studies.}  \textbf{(a)} In Fig.~\ref{fig:4} in the main text, macrodimers were excited in a resonant two-photon transition using a strong coupling field and a weak tunable field, realized by modulating sidebands on the coupling laser. \textbf{(b)} This scheme has also been used in our previous publication where macrodimer blockade has been observed, see Fig.~[12] in Ref.~\cite{Hollerith2023}. Surprisingly, the same dataset also showed correlated four-atom loss. Evaluating $G^{(2)}(\delta \mathbf{R})$ and $G^{(3)}_{\mathbf{R}_0}(\delta \mathbf{R})$ in this dataset shows the four-atom loss signal also in the three-atom loss correlations for $\delta \mathbf{R} = (2,-2)a_{\mathrm{lat}}$ and $\delta \mathbf{R} = (-3,3)a_{\mathrm{lat}}$. \textbf{(c)} 
 For a magnetic field $\mathbf{B}$ and a linear light polarization $\mathbfit{\varepsilon}$ pointing out of the atomic plane,  macrodimers can be excited along both lattice diagonal directions, as observed in the two-atom loss correlation signals $G^{(2)}(\mathbf{R}_0)$ and $G^{(2)}(\mathbf{R}_\perp)$, again using the phase-modulated laser. 
 Furthermore, $G^{(3)}_{\mathbf{R}_0}(-\mathbf{R}_0)$ and $G^{(3)}_{\mathbf{R}_\perp}(-\mathbf{R}_\perp)$ showed the three-atom loss signal along both directions. \textbf{(d)} The macrodimers studied here can also be excited using only a single strong laser field at a detuning $\mathrm\Delta_\mathrm{C} / 2\pi = -367.6\,\si{\mega\hertz}$ from the singly-excited states. 
   After a pulse time $70\,\si{\micro\second}$, we again observed correlated two-atom and three-atom losses. This excludes Rydberg antiblockade as an alternative mechanism behind the observed three-atom loss. The measurement operated in an intermediate regime between Fig.~\ref{fig:4}\,\textbf{(a)} and Fig.~\ref{fig:4}\,\textbf{(b)}, where dissipation is expected to be non-negligible but macrodimer hopping was strong enough to observe three-atom loss.}
  \end{figure*} 

In our system, Rydberg states $\ket{e}$ experienced a repulsive lattice potential whose magnitude is similar to the attractive potential experienced by ground state atoms $\ket{g}$.
Because the motional states coupled from the macrodimer states have kinetic energies larger than the on-site trapping frequency, we treated the atom pairs $\ket{ge}$ and $\ket{eg}$ as free particles and neglected the lattice potential. 
This is supported by our observations where the macrodimeron resonance positions presented in Fig.~\ref{fig:2} were independent of the lattice depth. 
The two macrodimeron resonances only got more pronounced at deeper lattices.
This was due to the larger motional state overlaps between $\Phi_g(R)$ and $\Phi_v(R)$ and lower correlated background losses~\cite{Simonelli_2016}.

The theoretical macrodimer decay rate is $\gamma \approx 1/20 \si{\micro\second}$~\cite{Sibalic2017,Hollerith_2021}, half of the decayed macrodimers are expected to end up in interacting Rydberg pair states $\ket{nS\,n^\prime P}$ or $\ket{nD\,n^\prime P}$ due to black-body transitions.
From the correlations observed after the short pulse time $t_1 = 500\,\si{\nano\second} \ll 1/\gamma$ in Fig.~\ref{fig:4}\,\textbf{(a)} and also the time-dependent study in Fig.~\ref{fig:SI_g3_time}, we infer that macrodimer decay does not contribute to the three-atom loss mechanism. 


\subsubsection{Correlation functions}

Here, we define the correlation functions used to obtain the microscopic signatures shown in the main text and the supplementary material. The two-atom loss correlations between empty sites at distance $\delta \mathbf{R}$ are evaluated via

\begin{align}\label{eq:g2}
& G^{(2)}(\delta \mathbf{R}) = \left(\langle \hat{h}_{\mathbf{R}^\prime + \delta \mathbf{R}}\hat{h}_{\mathbf{R}^\prime} \rangle - \langle \hat{h}_{\mathbf{R}^\prime + \delta \mathbf{R}}\rangle\langle\hat{h}_{\mathbf{R}^\prime} \rangle\right)_{\mathbf{R}^\prime}
\\
&
= \left( \biggl< \left(\hat{h}_{\mathbf{R}^\prime + \delta \mathbf{R}} - \langle \hat{h}_{\mathbf{R}^\prime + \delta \mathbf{R}}\rangle \right) \left(\hat{h}_{\mathbf{R}^\prime } - \langle \hat{h}_{\mathbf{R}^\prime}\rangle \right) \biggr> \right)_{\mathbf{R}^\prime}.  \notag
\end{align}
Here, $\left(\,.\,\right)_{\mathbf{R}^\prime}$ denotes averaging over all sites $\mathbf{R}^\prime$ in the lattice and $\langle \, . \, \rangle$ averaging over experimental realizations.
The projector $\hat{h}_{\mathbf{R}^\prime}$ evaluates to $1$~($0$) for an empty~(occupied) site at position $\mathbf{R}^\prime$.

The connected three-atom loss correlations are defined as
\begin{align}\label{eq:G3}
& \hspace{25pt} G^{(3)}_{\mathbf{R}_0}(\delta\mathbf{R}) =
\Biggl( \biggl< \left(\hat{h}_{\mathbf{R}^\prime } - \langle \hat{h}_{\mathbf{R}^\prime}\rangle \right)
\\
& \left(\hat{h}_{\mathbf{R}^\prime + \mathbf{R}_0} - \langle \hat{h}_{\mathbf{R}^\prime + \mathbf{R}_0}\rangle \right)
\left(\hat{h}_{\mathbf{R}^\prime + \delta \mathbf{R}} - \langle \hat{h}_{\mathbf{R}^\prime + \delta \mathbf{R}}\rangle \right) 
\biggr> \Biggl)_{\mathbf{R}^\prime},\notag
\end{align}
with conventions being identical as in Eq.~\ref{eq:g2}.
By subtracting the average loss signals at the different sites, only genuine three-atom loss leads to a signal. 
Correlations between three lattice positions depend on two relative distances. 
Here, we fix one distance to $\mathbf{R}_0 = (-1,1)a_{\mathrm{lat}}$ (as indicated by the subindex) and only vary the second distance $\delta\mathbf{R}$.
Plotting $G^{(3)}_{\mathbf{R}_0}(\delta\mathbf{R})$ contains the relevant correlation signal $G^{(3)}_{\mathbf{R}_0}(-\mathbf{R}_0)$ and $G^{(3)}_{\mathbf{R}_0}(2\mathbf{R}_0)$ where we expect to observe the macrotrimeron signal and also shows its significance relative to the background at other distances.
Distances where $\delta\mathbf{R}$ is zero or identical to $\mathbf{R}_0$ were excluded from the plots.

\subsection{Further microscopic studies}


Here, we add additional correlation measurements to support our interpretation of the laser-induced macrodimer exchange. Experimental details such as the detunings $\mathrm\Delta_\mathrm{C}$ and $\delta_\mathrm{p}$, and the relevant Rabi frequencies of all measurements are shown in table~\ref{table:1}. We also provide the estimated ratio $P_{2,3}$ of two-atom and three-atom losses. The ratio was extracted from numerically generated samples, where two-atom and three-atom loss rates were selected such that the simulated images reproduced the observed correlation strengths.

\subsubsection{Four-atom correlations} 
In addition to the previously discussed three-atom loss $G^{(3)}_{\mathbf{R}_0}(\delta \mathbf{R})$ at distances $\delta\mathbf{R}=(1,-1)a_{\mathrm{lat}}$ and $\delta\mathbf{R}=(-2,2)a_{\mathrm{lat}}$, some of the presented datasets also showed a signal at larger distances $\delta\mathbf{R}=(2,-2)a_{\mathrm{lat}}$ and $\delta\mathbf{R}=(-3,3)a_{\mathrm{lat}}$.
In our interpretation, this signal originates from a macrodimer delocalized over four sites due to the strong macrodimer hopping.
Because of the small overlap with the four-atom lattice ground state, which is even smaller than the overlap between the three-atom ground state and the macrotrimeron, the signal is small, but can still be observed.
The signal was particularly prominent in a statistically highly significant dataset which also showed macrodimer blockade, with experimental conditions very similar to Fig.~\ref{fig:4}\,\textbf{(a)}, see Fig.~\ref{fig:S1}\,\textbf{(b)}~\cite{Hollerith2023}.

\subsubsection{Orientation dependence}
We also excluded the laser propagation direction $\mathbf{k}_L$ to contribute to the process.
We therefore rotated the magnetic field and the UV polarization out of the atomic plane such that macrodimers were excited along both lattice diagonal directions $\mathbf{R}_0=(-1,1)\,a_{\mathrm{lat}}$ and $\mathbf{R}_\perp=(-1,-1)\,a_{\mathrm{lat}}$, see Fig.~\ref{fig:S1}\,\textbf{(c)}. 
After an illumination time $t_{3} = 4\,  \si{\micro\second}$ on two-photon resonance, we observed similar two-atom and three-atom loss correlations along the diagonal direction parallel and orthogonal to $\mathbf{k}_L$. Here, in addition to the three-atom loss correlations $G^{(3)}_{\mathbf{R}_0}(\delta\mathbf{R})$ defined in Eq.~\ref{eq:G3}, we evaluated three-atom loss correlations $G^{(3)}_{\mathbf{R}_\perp}(\delta\mathbf{R})$, which quantify the loss of a third atom conditioned on an atom pair lost at a distance $\mathbf{R}_\perp$.

\subsubsection{Detuning dependence}

Next, we switched the sideband modulation off and excited macrodimers using a single strong two-photon resonant light field~\cite{Hollerith_2019,Hollerith_2021}, probing the regime of large detunings $\mathrm\Delta_\mathrm{C}/2\pi \approx U/2$ between the macrodimer and the singly-excited states, see Fig.~\ref{fig:S1}\,\textbf{(d)}. 
We again observed significant three-atom loss. 
This excludes Rydberg antiblockade, where the presence of a macrodimer facilitates secondary Rydberg excitations of nearby atoms due to the contributing interaction shifts~\cite{Simonelli_2016,Goldschmidt_2016}, because antiblockade sensitively depends on the laser frequency which is far-detuned from the previous configuration using the near-resonant sideband:
At different laser frequencies, antiblockade is expected to occur at different distances and/or orientations.
However, it does not exclude our light-induced hopping process, which now has a reduced coupling rate but is still two-photon resonant.
For most of the measurements for small $\mathrm\Delta_\mathrm{C}$ and $\mathrm\Omega_\mathrm{C}/2\pi  \gtrsim 1 \si{\mega\hertz}$, the coupling between macrodimers was dominant.
For an intermediate-state detuning $\mathrm\Delta_\mathrm{C}/2\pi = -367.6\,\si{\mega\hertz}$ and $\mathrm\Omega_\mathrm{C}/2\pi = 2.9(3)\,\si{\mega\hertz}$, two-photon macrodimer hopping rate can be approximated via $\mathrm\Omega^{(2)}_\mathrm{h} = \frac{\mathrm\Omega_\mathrm{C}^2}{2 \mathrm\Delta_\mathrm{C}} \widetilde f \approx 2\pi \times  4\,\si{\kilo\hertz} $, with $\widetilde f \approx 0.35$ the estimated motional state overlap.
The observed correlation signal (see table~\ref{table:1}) is plausible accounting for the fact that excited macrodimers have only a fnitite lifetime and are furthermore expected to loose the spatial overlap with the lattice after several microseconds because they are released from the optical trap.
\begin{figure}
  \centering
  \includegraphics[width=1.0\columnwidth]{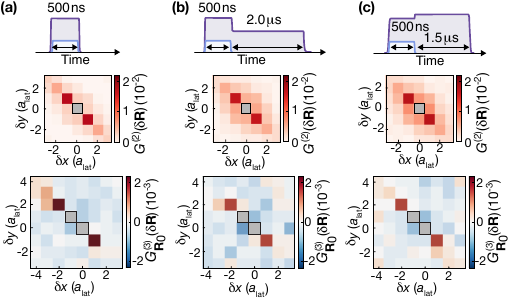}%
  \caption{\label{fig:SI_g3_time}\textbf{Dynamics after a macrodimer seed pulse.} \textbf{(a)} Signal after the $t_1 = 500\,\si{\nano\second}$ UV pulse, also shown in Fig.~\ref{fig:4}\,\textbf{(a)}. 
  \textbf{(b)} Switching the modulated sideband off but keeping the strong and off-resonant carrier field on for additional $t_C = 2\,\si{\micro\second}$, with the overall UV power reduced to half of its initial value, we observed no further increase of the signals $G^{(2)}(\mathbf{R}_0)$ and $G^{(3)}_{\mathbf{R}_0}(-\mathbf{R}_0)$. Furthermore, we observed an increased background in atom pair-loss correlations at other distances from excitations facilitated by the seeded Rydberg excitations. \textbf{(c)} The background became even more pronounced when keeping the UV laser at full power for $t_C = 1.5\,\si{\micro\second}$.  These observations are inconsistent with alternative interpretations of the three-atom correlation signal which rely on Rydberg decay, where keeping the carrier field on after the pulse time $t_1$ is expected to increase the three-atom loss signal.} 
  \end{figure}

   \begin{table*}
\begin{tabular}{c c c c c c c c c c}
\\[-2ex]\hline 
     \hline \\[-2ex] 
 Dataset & $t$~($\si{\micro\second}$) & $\mathrm\Omega_\mathrm{C}/2\pi$~(MHz)   & $\mathrm\Delta_\mathrm{C}/2\pi$~(MHz) & \hspace{1mm} $\delta_\mathrm{p}/2\pi$~(MHz) & \hspace{1mm} $\rho_f$ \hspace{1mm} & \hspace{3mm} $G^{(2)}(\mathbf{R}_0)$ \hspace{1mm} & \hspace{1mm} $G^{(3)}_{\mathbf{R}_0}(-\mathbf{R}_0)$   & $P_{2,3}$ & Images\vspace{2pt} \\  \hline
 Fig.~\ref{fig:3}\,\textbf{(c)}  & 50  & 5.7(2)\, & -2.6  & -2.6  &  $78\%$  & $2.40(5)\times 10^{-2}$  & $4.8(3)\times 10^{-3}$ & 3:1 & 927  \\
 Fig.~\ref{fig:4}\,\textbf{(a)}\,$\&$\,\ref{fig:SI_g3_time}\,\textbf{(a)}  & 0.5 & 7.8(3) & -5  & 5  &  $75\%$ & $2.29(3)\times 10^{-2}$ & $2.1(3)\times 10^{-3}$ & 5:1 & 1358 \\
 Fig.~\ref{fig:4}\,\textbf{(b)}   & 500 & low & -3.6  & 3.6  &  $82\%$ &  $2.32(3)\times 10^{-2}$ & $-1.1(2)\times 10^{-3}$ & large & 3435 \\
    Fig.~\ref{fig:S1}\,\textbf{(b)}  & 2 & 2.6(3) & -3.6  & 3.6 & $86\%$ & $1.65(1)\times 10^{-2}$ & $0.97(6)\times 10^{-3}$ & 10:1 & 21095 \\
  Fig.~\ref{fig:S1}\,\textbf{(c)}  &4  & 1.7(5) & -3.6  & 3.6  &  $80\%$ &  $1.55(2)\times 10^{-2}$ & $0.67(10)\times 10^{-3}$ & 15:1 & 9433 \\
      Fig.~\ref{fig:S1}\,\textbf{(d)}    & 70 & 2.9(3) & -367  & 367  &  $84\%$ &  $2.70(3)\times 10^{-2}$ & $1.4(2)\times 10^{-3}$ & 10:1 & 4022 \\
   Fig.~\ref{fig:SI_g3_time}\,\textbf{(b)}  & 0.5; 2 & 7.8(3); 6.2(2)  & -5  & 5  &  $69\%$ & $2.39(4)\times 10^{-2}$& $1.5(2)\times 10^{-3}$  & 7:1 & 1335 \\
 Fig.~\ref{fig:SI_g3_time}\,\textbf{(c)}  & 0.5; 1.5 & 7.8(3); 8.8(3)  & -5  & 5  & $67\%$ & $2.50(4)\times 10^{-2}$& $1.5(2)\times 10^{-3}$  & 7:1 & 1415 \\
\\[-2ex]\hline 
     \hline \\[-2ex] 
\end{tabular}
\caption{\textbf{Further experimental details.} Illumination times $t$, Rabi frequency $\mathrm\Omega_\mathrm{C}$, detunings $\mathrm\Delta_\mathrm{C}$ and $\delta_\mathrm{p}$, final densities $\rho_f$ after the UV pulse, correlations $G^{(2)}(\mathbf{R}_0)$ and $G^{(3)}_{\mathbf{R}_0}(-\mathbf{R}_0)$, the estimated ratio $P_{2,3}$ between two-atom and three-atom events in order to get similar correlation signals when sampling from numerically generated images, and the number of experimental shots. The negative $G^{(3)}_{\mathbf{R}_0}(-\mathbf{R}_0)$ signal for low $\mathrm\Omega_\mathrm{C}$ can be understood from pair losses alone and was also found in our simulated images: Atoms at a distance $\mathbf{R}_0$ from a previously excited macrodimer have one less neighbor to form another subsequent macrodimer excitation, leading to a negative correlation signal. For the dataset presented in Fig.~\ref{fig:S1}\,\textbf{(c)} where macrodimers were excited along two directions, the correlation strengths perpendicular to $\mathbf{R}_0$ were $G^{(2)}(\mathbf{R}_\perp) = 1.30(2)\times 10^{-2}$ and $G^{(3)}_{\mathbf{R}_\perp}(-\mathbf{R}_\perp) = 0.75(10)\times 10^{-3}$. Correlations were analyzed in a region-of-interest of $11\times 11$ or $15\times 15$ lattice sites in the center of the lattice. Statistical errors on the correlation signals were calculated using a bootstrap algorithm (delete-1 jackknife).} 
\label{table:1}
 \end{table*}

\subsubsection{Time-dependent correlation signal}
Finally, we studied the dynamics of the signal after a macrodimer seed.
After the $t_1 = 500\,\si{\nano\second}$ pulse (see also Fig.~\ref{fig:4}\,\textbf{(a)}), we switched the sideband and therefore macrodimer excitation from the ground state off but kept the carrier field on and studied the further dynamics of the correlation signal, see Fig.~\ref{fig:SI_g3_time}. 
We did not observe a further increase of the three-atom loss correlations. 
This is consistent with our expectation since the time associated with the macrodimer hopping is larger than the time of the seed pulse $t_1$, suggesting excitations which have a large overlap with macrodimerons and macrotrimerons as stable eigenstates of the coupled light-matter Hamiltonian, instead of isolated macrodimers which then undergo hopping dynamics.  Furthermore, the observations are inconsistent with alternative mechanisms based on Rydberg decay where an increase of the correlation signals would be expected.

In the same datasets, we also observed a slight decrease of the overall density and an increase of the two-atom loss correlation background between nearby sites (not only at distance $\mathbf{R}_0$).
In our interpretation, the background correlations originated from secondary Rydberg excitations created by our $\Delta^{ge}_\mathrm{C}/2\pi \approx 730\,\si{\mega\hertz}$ detuned carrier field and were facilitated by the macrodimer seed pulse~\cite{Simonelli_2016}. 
Without the initial seed pulse, no dynamics was observed on this short timescale.

\subsection{Further spectroscopic studies}
  \begin{figure}
  \centering
  \includegraphics[width=1.0\columnwidth]{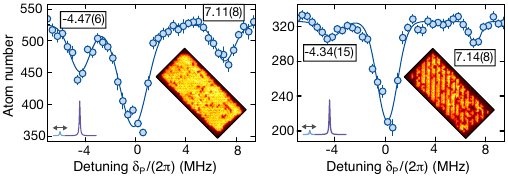}%
  \caption{\label{fig:SI_S4}\textbf{Dependence on the atom arrangement.} We compared the spectrum for $\mathrm\Omega_\mathrm{C}/2\pi = 6.2\,\si{\mega\hertz}$ in Fig.~\ref{fig:2}\,\textbf{(a)} measured in an unity-filled atomic Mott insulator with the signal obtained from an arrangement where every third lattice row was unoccupied and three-atom contributions were absent (as shown in the exemplary images of the prepared atom configurations). Again, the sideband frequency was varied while the carrier field was kept unchanged (see left insets). The uncertainy of the fitted resonance position (insets) was similar as the observed shift between both configurations.
  All error bars on data points represent one s.e.m., errors on the resonance position represent estimated fit errors.}
  \end{figure} 

Here, we present three additional atom loss spectra showing the presence of macrodimerons.

First, we tried to spectroscopically resolve the small difference expected from the calculated two-atom spectra and the three-atom spectra, see Fig.~\ref{fig:2}\,\textbf{(c)} and Fig.~\ref{fig:3}\,\textbf{(b)} in the main text.
We therefore initialized an atomic arrangement where every third lattice row was unoccupied and compare the spectrum with the one observed in the unity-filled lattice. 
For the density-modulated arrangement where macrodimer exchange cannot occur and macrotrimeron contributions are absent, all physics should be captured by the two-atom Fano model introduced in Eq.~\ref{eq:Fano_H}.
We again measured atom loss for varying modulation frequencies while keeping the carrier frequency constant.
To avoid systematic shifts from drifting experimental conditions, the images of both initial arrangements were taken alternately.
Because the observed shift for the resonance at $\delta_\mathrm{p}<0$ between both configurations is as large as the uncertainty of the fit, we cannot claim to resolve the macrotrimeron spectroscopically.
We found that the two resonances shifted by the strong coupling field are more pronounced in the Mott insulating state.
We attribute this to the ratio between the available macrodimers and singly-excited Rydberg atoms, which is twice as large in the Mott insulating state.
\begin{figure}
  \centering
  \includegraphics[width=1.0\columnwidth]{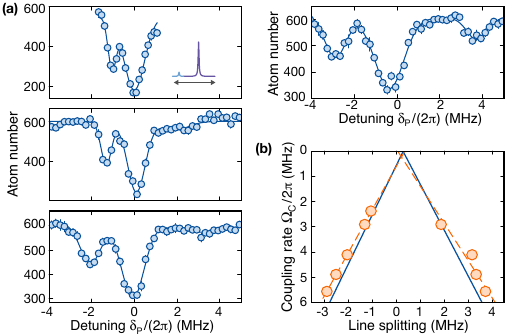}%
  \caption{\label{fig:SI_S5}\textbf{Spectroscopy for fixed sideband frequency.}  \textbf{(a)} In contrast to Fig.~\ref{fig:2}\,\textbf{(a)} where only the sideband frequency was varied, the sideband frequency was set equal to the interaction shift $U$ and the overall laser frequency was varied (as indicated in the inset). In this configuration where both detunings $\delta_\mathrm{p} = \mathrm\Delta_\mathrm{C}$ are equal, all resonances are slightly closer. 
  The spectra were recorded at coupling rates $\mathrm\Omega_\mathrm{C}/2\pi = \left[2.5,4.3,5.1,5.8\right]\,\si{\mega\hertz}$. 
   The correlations shown in Fig.~\ref{fig:3}\,\textbf{(c)} were measured at the left peak for $\mathrm\Omega_\mathrm{C} = 5.8\,\si{\mega\hertz}$ with $\delta_\mathrm{p}/2\pi = -2.6\,\si{\mega\hertz}$. \textbf{(b)} As one can see from the fit (orange dashed line), the splitting again depends linearly on $\mathrm\Omega_\mathrm{C}$. For this dataset, the theoretical prediction (solid blue line) agrees well with the observations. Error bars on data points represent one s.e.m. }
  \end{figure} 

Second, we performed a reference spectroscopy starting from the Mott insulator where we keep the modulation frequency at the interaction shift $U$ and vary the overall laser frequency instead, see Fig.~\ref{fig:SI_S5}. 
 Here, if the probe field is on resonance with the $\ket{g}\rightarrow \ket{e}$ transition  transition where $\delta_\mathrm{p} = 0$, also the carrier field is resonant with the $\ket{ge},\ket{eg} \rightarrow \ket{\Psi_v}$ transition where $\mathrm\Delta_\mathrm{C}=0$.
Now, varying the overall laser frequency during the spectroscopy tunes both detunings $\delta_\mathrm{p} = \mathrm\Delta_\mathrm{C}$ equally.
 The spectra look qualitatively similar as in Fig.~\ref{fig:2}\,\textbf{(a)}. The only difference is that the resonances appear at slightly lower detunings $|\delta_\mathrm{p}|$.

       \begin{figure}
  \centering
  \includegraphics[width=1.0\columnwidth]{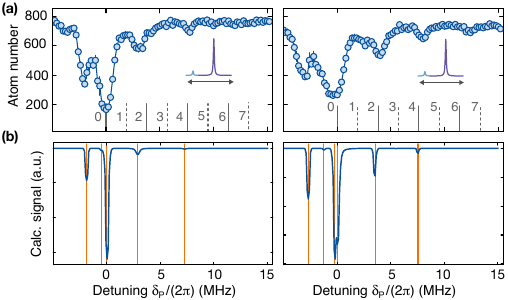}%
  \caption{\label{fig:SI_S3}\textbf{Spectra including higher vibrational modes.} \textbf{(a)} Additional spectra for Rabi couplings $\mathrm\Omega_\mathrm{C}/2\pi = 4.1(2)\,\si{\mega\hertz}$ (left) and $\mathrm\Omega_\mathrm{C}/2\pi = 5.7(2)\,\si{\mega\hertz}$ (right) including a wider detuning range. As in Fig.~\ref{fig:SI_S5}, the overall UV laser frequency was varied and the phase modulation frequency remained fixed and equal to the interaction shift $U$ (see inset). The solid (dashed) gray lines indicates the theoretical position of the even (odd) vibrational macrodimer levels in the absence of light shifts. All error bars on data points represent one s.e.m. \textbf{(b)} The corresponding calculated  two-atom spectra obtained when including the six lowest vibrational states. The Rabi frequencies were identical as in the pictures above. While macrodimerons corresponding to the lowest vibrational mode are still dominating the spectrum, also signals of higher modes appear. Their contribution explains the observed broadening of the central line at $\delta_\mathrm{p} \approx 0$ towards negative detunings, see also Fig.~\ref{fig:Fano_phase_theory} as well as Fig.~\ref{fig:2}\,\textbf{(a)} in the main text and Fig.~\ref{fig:SI_S5}\,\textbf{(a)}.}
  \end{figure} 

The discussion in the main text neglected contributions of higher vibrational macrodimer modes.
 In a last measurement, we performed spectroscopy over a wider frequency range where also higher modes became resonant with the laser light, see Fig.~\ref{fig:SI_S3}\,\textbf{(a)}.
 As in the previous paragraph and Fig.~\ref{fig:SI_S5}, the overall laser frequency was varied. 
Due to the spatial overlap with the lattice ground state, the lowest vibrational mode still contributes the most. 
 In the corresponding calculations shown in Fig.~\ref{fig:SI_S3}\,\textbf{(b)}, different vibrational states remain uncoupled and hybridize individually into macrodimerons because of their orthogonality. Accounting also for the $R-$dependency of the optical coupling discussed in Fig.~\ref{fig_pot_pert} in principle modifies the vibrational modes in the binding potential and enables coupling between different vibrational states.

\newpage
\end{document}